\documentclass[%
preprint,
superscriptaddress,
citeautoscript,
 amsmath,amssymb,
prb,
floatfix,
]{revtex4-1}
\usepackage[version=3]{mhchem}
\usepackage{graphicx}
\usepackage{dcolumn}
\usepackage{bm}
\usepackage{multirow}
\usepackage{booktabs}
\usepackage{multirow}
\usepackage{siunitx}
\usepackage{xcolor}

\usepackage{pdfpages}
\usepackage{etoolbox} 

\makeatletter
\patchcmd{\@outputpage@head}{\@ifx{\LS@rot\@undefined}{}{\LS@rot}}{}{}{}
\makeatother

\begin{document}

\preprint{\textcolor{blue}{This is a post-peer-review, pre-copyedit version of an article published in \textit{Nature Materials}.}}
\preprint{\textcolor{blue}{The final authenticated version is available online at: \url{https://doi.org/10.1038/s41563-018-0262-7.}}}

\title{Phonon coherences reveal the polaronic character of excitons in two-dimensional lead-halide perovskites}

%

\author{F\'elix~Thouin}
\affiliation{School of Physics, Georgia Institute of Technology, 837 State Street NW, Atlanta, Georgia 30332, USA}

\author{David~A.~Valverde-Ch\'avez}%
\affiliation{School of Chemistry and Biochemistry, Georgia Institute of Technology, 901 Atlantic Drive NW, Atlanta, Georgia 30332, USA}

\author{Claudio~Quarti}
\affiliation{Laboratory for Chemistry of Novel Materials, Department of Chemistry, Universit\'e de Mons, Place du Parc 20, 7000, Mons, Belgium}

\author{Daniele~Cortecchia}
\affiliation{Center for Nano Science and Technology@PoliMi, Istituto Italiano di Tecnologia, via Giovanni Pascoli 70/3, 20133 Milano, Italy}

\author{Ilaria~Bargigia}%
\affiliation{School of Chemistry and Biochemistry, Georgia Institute of Technology, 901 Atlantic Drive NW, Atlanta, Georgia 30332, USA}

\author{David~Beljonne}
\affiliation{Laboratory for Chemistry of Novel Materials, Department of Chemistry, Universit\'e de Mons, Place du Parc 20, 7000, Mons, Belgium}

\author{Annamaria~Petrozza}
\affiliation{Center for Nano Science and Technology@PoliMi, Istituto Italiano di Tecnologia, via Giovanni Pascoli 70/3, 20133 Milano, Italy}

\author{Carlos~Silva}%
\affiliation{School of Chemistry and Biochemistry, Georgia Institute of Technology, 901 Atlantic Drive NW, Atlanta, Georgia 30332, USA}
\affiliation{School of Physics, Georgia Institute of Technology, 837 State Street NW, Atlanta, Georgia 30332, USA}
\email{carlos.silva@gatech.edu}

\author{Ajay~Ram~Srimath~Kandada}%
\affiliation{School of Physics, Georgia Institute of Technology, 837 State Street NW, Atlanta, Georgia 30332, USA}
\affiliation{School of Chemistry and Biochemistry, Georgia Institute of Technology, 901 Atlantic Drive NW, Atlanta, Georgia 30332, USA}
\affiliation{Center for Nano Science and Technology@PoliMi, Istituto Italiano di Tecnologia, via Giovanni Pascoli 70/3, 20133 Milano, Italy}
\email{srinivasa.srimath@iit.it}


\date{\today}

\begin{abstract}
\textbf{
Hybrid organic-inorganic semiconductors feature complex lattice dynamics due to the ionic character of the crystal and the softness arising from non-covalent bonds between molecular moieties and the inorganic network. Here we establish that such dynamic structural complexity in a prototypical two-dimensional lead iodide perovskite gives rise to the coexistence of diverse excitonic resonances, each with a distinct degree of polaronic character. By means of high-resolution resonant impulsive stimulated Raman spectroscopy, we identify vibrational wavepacket dynamics that evolve along different configurational coordinates for distinct excitons and photocarriers. Employing density functional theory calculations, we assign the observed coherent vibrational modes to various low-frequency ($\lesssim 50$\,cm$^{-1}$) optical phonons involving motion in the lead-iodide layers. We thus conclude that different excitons induce specific lattice reorganizations, 
which are signatures of polaronic binding. This insight on the energetic/configurational landscape involving globally neutral primary photoexcitations may be relevant to a broader class of emerging hybrid semiconductor materials.  }
\end{abstract}

\maketitle

Hybrid organic-inorganic metal-halide perovskite quantum-well-like derivatives are of increasingly sharp focus due to the presence of strongly bound, stable excitons at room temperature~\cite{Ishihara1989,Even2014,Even2015,Saparov2016,blancon2018,Neutzner2018}. These excitons are viewed broadly as analogous to those in epitaxial semiconductor quantum wells, which generally feature much lower binding energies. Nevertheless, the distinct ionic character, and the `softness' of the lattice resulting from organic counter-ion coordination, give rise to strong electron-phonon coupling and dynamic disorder effects, which influence the optical and electronic properties of this class of materials~\cite{Yaffe2015,kandadad2016photophysics,miyata2017large}, and their consequences on excitonic structure are not thoroughly explored.  There are therefore currently open questions on the peculiar nature of excitons in these structurally complex materials, which are argued to be in an intermediate regime between extended Wannier excitons in quantum-confined semiconductors and localized excitons in molecular semiconductors~\cite{straus2018_2}. We establish here that excitons in two-dimensional (2D) hybrid perovskites are dressed by the ionic lattice leading to the coexistence of multiple excitons with distinct lattice couplings. This has consequences in the formation of biexcitons~\cite{Thouin2017}, for example, which would have profound implications for the development of light-emitting devices~\cite{Kondo1998,Quan:2017aa,Su:2017aa,Booker2018}. More generally, our conclusions will shape the detailed description of fundamental excitonic processes --- energy transport, population dynamics, quantum dynamics (involving dynamics of many-body couplings and of dephasing dynamics, for example) in 2D hybrid perovskites. Beyond these materials, this knowledge represents an acute contribution to the comprehension of ionic semiconductors with elaborate hybrid lattices~\cite{Senger2003aa,Dvorak2013aa}. 

Unlike excitons in (non-ionic) semiconductor quantum wells, for which their resonances are characterized by sharp, structureless spectral lineshapes, excitons in these materials display rich spectral structure that depends on the degree of lattice distortion imposed by the organic cationic ligands~\cite{Ishihara1990,kataoka1993magneto,Tanaka2005,Shimizu2005,Ema2006,Goto2006,Kitazawa2010a,Gauthron:10,Straus2016a,quarti2018tuning,Neutzner2018}.  
To account for this structure, we have invoked a general framework of a delocalized Wannier exciton with binding energy $\sim 200$\,meV but with substantially large coupling to local lattice vibrations,  
modelled by four distinct, non-degenerate excitonic transitions, spectrally separated by multiples of  
35\,meV~\cite{Neutzner2018}. 
Previously, we hypothesized that polaronic effects play a role 
in such a rich excitonic spectral structure~\cite{Neutzner2018} ubiquitously observed in 2D perovskites. A direct optical probe of polaronic effects on the nature of excitons would be resonance-Raman spectroscopy~\cite{sood1985resonance}. However, the high photoluminescence background in these materials obscures the relatively weak Raman signal, making such a measurement a practically arduous task. Here, we find direct and unambiguous evidence for this hypothesis by implementing  
high-resolution resonant impulsive stimulated Raman spectroscopy (RISRS)~\cite{dhar1994time,merlin1997generating}. We establish  polaronic effects on discrete non-degenerate excitonic transitions, evident via distinct coupling to low-frequency phonons. We find that the coupling of carriers to lattice degrees of freedom is stronger than that for excitons in general, which we rationalize by the ionic nature of the crystal. Nevertheless, different excitons with common ground state and with specific polaronic character coexist in this class of materials.

\begin{figure}[ht]
\centering
\includegraphics[width =0.95\textwidth]{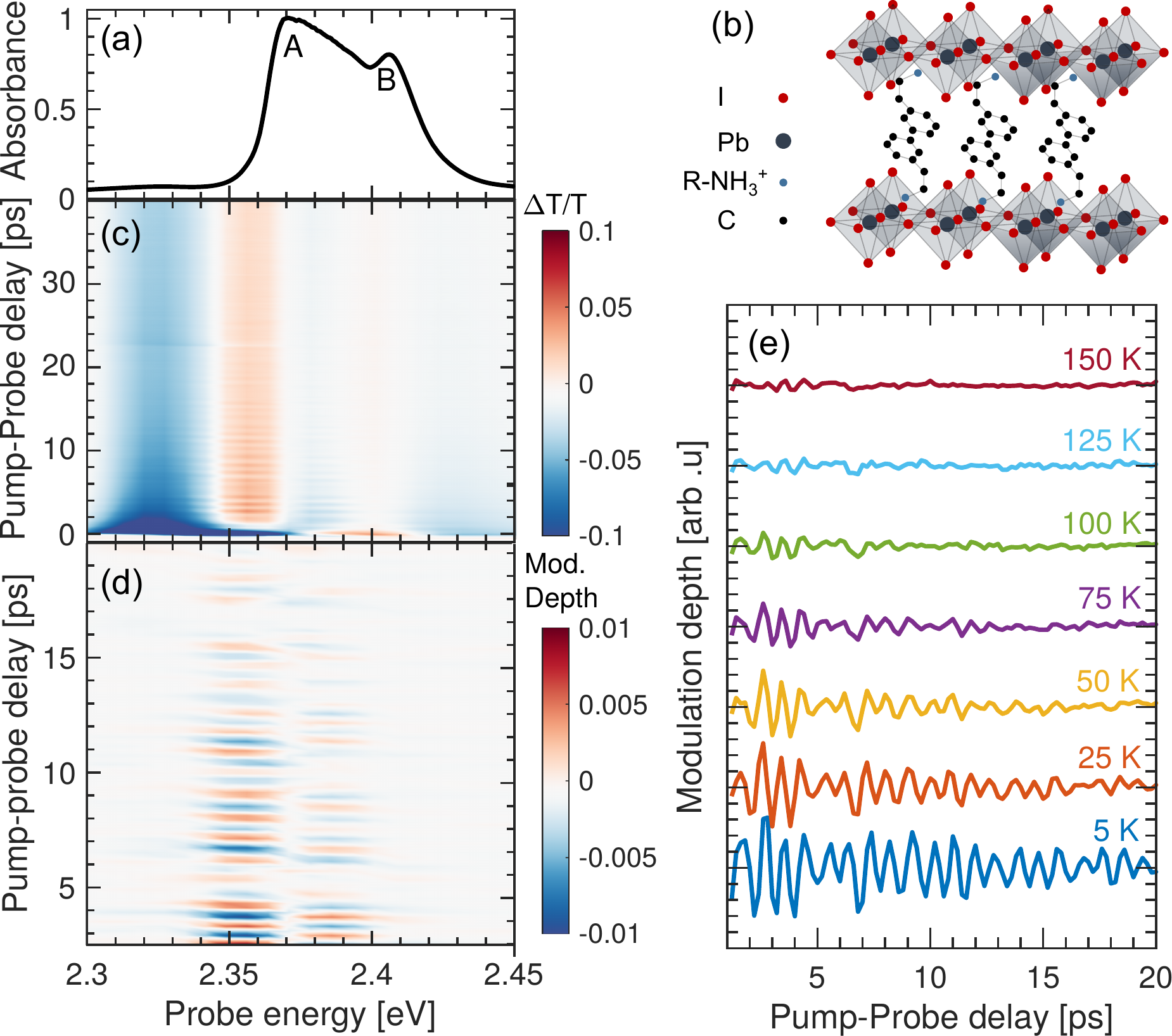}
\caption{\textbf{Impulsive coherent vibrational dynamics of \ce{(PEA)2PbI4}.} (a) Absorption spectrum measured at 5\,K. (b) Schematic of the lattice structure of \ce{(PEA)2PbI4} (see Supporting Information section S1 for X-Ray diffraction data). (c) Time-resolved differential transmission spectrum measured at 5\,K and with pump photon energy of $3.06$\,eV. (d) Oscillatory components of the time-resolved differential transmission spectrum, obtained by subtracting the population dynamics (empirically approximated as twelfth-order polynomials) from the transient spectrum in part (c). (e) Temperature-dependent cuts of the oscillatory response obtained by binning transient maps such as that displayed in part (d) around a probe energy of $2.35$\,eV.}
\label{fig1}
\end{figure}

We focus on two prototypical single-layer perovskite systems, \ce{(PEA)2PbI4} (PEA = phenylethylammonium) and \ce{(NBT)2PbI4} (NBT = n-butylammonium), which have slightly different structural disortions imposed by the organic cation~\cite{Cortecchia2017a}.  
The exciton absorption spectrum of \ce{(PEA)2PbI4} 
measured at 5\,K, is shown in Fig.~\ref{fig1}(a) and the representative crystal structure is depicted in Fig.~\ref{fig1}(b). The  
spectrum is characterized by a very well defined  
lineshape composed of two dominant transitions at 2.37 and 2.41\,eV, labeled A and B, respectively in Fig.~\ref{fig1}(a), and two additional peaks with the same energy spacing above and below these two main peaks, with much lower oscillator strength~\cite{Neutzner2018}. We have carried out transient absorption measurements by pumping into the conduction band at 3.06\,eV,  
shown in Fig.~\ref{fig1}(c). These excitation conditions generate an initial hot charge-carrier density.  
The spectra and dynamics follow the reported trends dominated by carrier thermalization and exciton-screening mechanisms~\cite{guo2016electron,grancini2015role}. During the first picosecond, the differential transmission spectrum is composed of a strong negative feature across the A band and a positive feature at the B band. This lineshape is characteristic of a superposition of excitation-induced shift and broadening of the excitonic transition by many-body interactions along with the bleach of higher lying electronic states~\cite{Haug2008}. In a few picoseconds, the positive signal corresponding to the ground-state bleach gains in intensity, indicating carrier thermalization into the A exciton. The spectrum still contains the signatures of pump-induced spectral shifts induced by the substantial photocarrier population over picosecond time windows. The low-energy negative feature below 2.3\,eV can be attributed to the excited-state absorption from the exciton to mutli-particle states~\cite{Thouin2017}.  

In addition to the population dynamics, we observe a periodic modulation of the differential transmission signal, particularly strong in the spectral region that corresponds to absorption of exciton A. The oscillatory components can be clearly seen after subtracting the population dynamics at all detection energies as shown in Fig.~\ref{fig1}(d). We identify these as the signatures of coherent phonons generated via resonant impulsive stimulated Raman scattering (RISRS) induced by the ultrashort pump pulse~\cite{dhar1994time}. When the duration of the pulse is much shorter than the period of Raman-active low-frequency vibrations, Raman interactions generate an impulsive force on the lattice driving its coherent motion~\cite{dhar1994time}. This modulates the permittivity at the frequency of the lattice motion, which can be detected as the oscillatory component of the differential transmission signal.  

As shown in Fig.~\ref{fig1}(e), at higher temperatures the coherent oscillations not only dephase faster due to phonon-phonon scattering~\cite{ivanovska2016vibrational}, but they also exhibit reduced modulation depth. The latter can be attributed to the presence of strong dynamic disorder, especially above 100\,K as we have demonstrated previously~\cite{Thouin2017}. We only focus on the coherent phonon dynamics at 5\,K in the rest of the manuscript because the conclusions that we draw by analyzing the vibrational coherences under resonant excitation at low temperature are relevant over the entire range up to room temperature, as we have established that the same excitonic spectral structure persists over this range~\cite{Thouin2017,Neutzner2018}.  The full temperature-dependent dataset is presented in Figs. S2 and S3 of SI.

\begin{figure}[ht]
\centering
\includegraphics[width =1.0\textwidth]{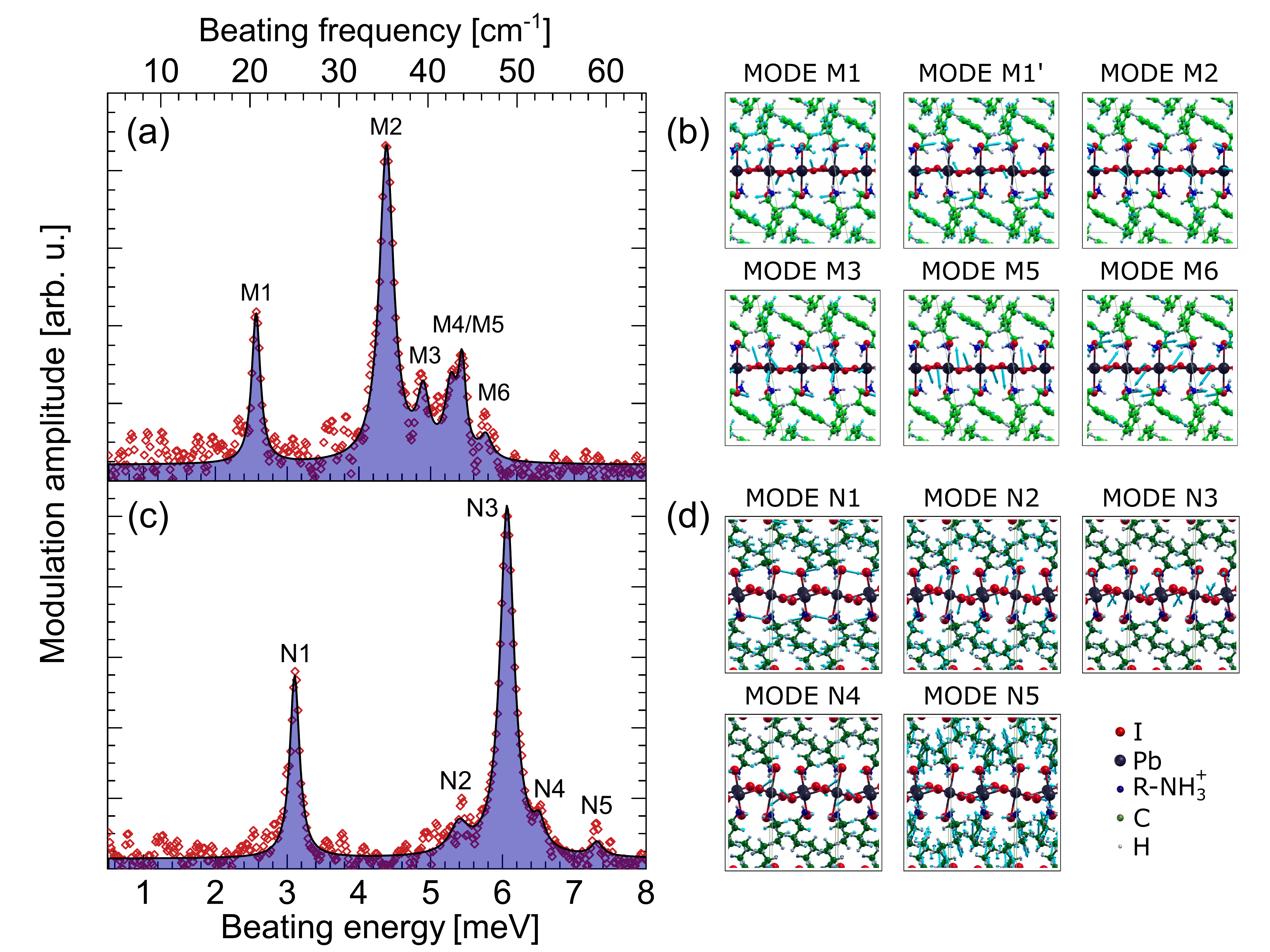}%
\caption{\textbf{Integrated resonant impulsive stimulated Raman spectrum of \ce{(PEA)2PbI4} and \ce{(NBT)2PbI4} at 5\,K and associated phonon modes.} (a) and (c) Fourier-transform spectrum of the oscillatory components such as those displayed in Fig.~\ref{fig1}(d), and integrated over all the detection energies probed in Fig.~\ref{fig1} for \ce{(PEA)2PbI4} and \ce{(NBT)2PbI4} respectively. The pump energy is tuned to 3.06\,eV and 3.1\,eV respectively. The continuous line with blue shade is a fit to a function composed of six and five Lorentzian components respectively. (b) and (d) Diagrammatic representations of the vibrational modes obtained via DFT calculations. The black spheres at the center of the octahedra represent the Pb ions, while the red spheres represent the iodine. The green structures represent the organic cations - PEA in (b) and NBT in (d). The cyan arrows represent the motion of the each of the ions within the lattice.}
\label{fig_spec}
\end{figure}

By Fourier-transforming the measured oscillatory response in Fig.~\ref{fig1} along the pump-probe time axis, we obtain a RISRS spectrum. We identify six vibrational modes that we label M1--M6, with peak energies reported in Table~\ref{Table}. The peaks display well-defined Lorentzian lineshapes with full-width-at-half-maximum $\leq 0.33$\,meV, as shown by the fit in Fig.~\ref{fig_spec}(a). 

\begin{table}[ht]
\centering
\caption{\textbf{Assignment of the resonant impulsive stimulated Raman spectrum of \ce{(PEA)2PbI4} and \ce{(NBT)2PbI4}.} 
Experimental vibrational energies from spectrum in Fig.~\ref{fig_spec}, and normal-mode energies obtained from density functional theory calculations, along with the mode assignment.}
  \begin{tabular}{ccccc}
      \hline
    \multirow{2}{*}{Mode} & Measurement & Calculation & $\lambda$ &
    \multirow{2}{*}{Mode Assignment}  \\ 
     & {[meV]}& {[meV]} & {[meV]} 
     \\
     \hline
    \multicolumn{5}{c}{\ce{(PEA)2PbI4}} \\
    \hline
     \midrule
   	  M1 & 2.57 $\pm$ 0.01 & 3.18 & 0.609 & Octahedral twist along a axis on the inorganic sheet \\
    M1$^\prime$ & {---}  & 3.93 & 0.94 & Octahedral twist along a axis on the inorganic sheet\footnote{Mode M1$^{\prime}$ is predicted by calculation  but is not discernible experimentally.}  \\
    M2 & 4.38 $\pm$ 0.01 & 4.51 & 5.21 & Octahedral twist and Pb-I-Pb bending \\
    M3 & 4.89 $\pm$ 0.02 & 4.51 & 3.49 & Pb Displacement and Pb-I-Pb bending  \\
    M4 & 5.22 $\pm$ 0.06 & {---} & {---} & {---}\footnote{Mode M4 is measured experimentally but not predicted by calculation.}\\
     M5 & 5.41 $\pm$ 0.02 & 5.34 & 4.54 & Pb-I-Pb bending and Pb-I stretching\\
    M6 & 5.75 $\pm$ 0.01 & 5.86 & 4.78 & Scissoring of Pb-I-Pb angle \\
   \hline 

    \multicolumn{5}{c}{\ce{(NBT)2PbI4}} \\
    \hline
    \midrule
   	  N1 & 3.10 $\pm$ 0.01 & 2.93 & 0.76 & Octahedral twist along a axis on the inorganic sheet \\
    N2 & 5.42 $\pm$ 0.02 & 5.38 & 4.42 & Octahedral twist along a axis on the inorganic sheet \\
    N3 & 6.06 $\pm$ 0.01 & 5.97 & 16.29 & Octahedral twist along axis orthogonal to the inorganic sheet  \\
    N4 & 6.52 $\pm$ 0.15 & 6.60 & 8.69 & Scissoring of Pb-I-Pb angle\\
     N5 & 7.31 $\pm$ 0.19 & 7.37 & 0.12 & Scissoring of Pb-I-Pb angle\\
   \hline 
  \end{tabular}
  \label{Table}

\end{table}

To properly discern the nature of coupling and to assign the observed energies to specific lattice vibrations, we calculated the vibrational normal modes of \ce{(PEA)2PbI4} by using density functional theory (DFT) within the harmonic approximation. This (almost) parameter-free computational approach represents the current state-of-the art in the simulation of the vibrational spectroscopic response of solid-state materials including hybrid perovskites~\cite{corno2006periodic, brivio2015lattice, quarti2013raman, ivanovska2016vibrational}. The calculations do not yield negative vibrational frequencies at the $\Gamma$ point, further indicating that the crystalline structure used to model \ce{PEA2PbI4} is a real minimum of the potential energy surface (the full list of computed vibrational frequencies is reported in Section S5 in SI). 

The electron-phonon coupling is estimated by displacing the crystalline structure along the normal mode ($Q_i$) and evaluating the corresponding variation of the single-particle electronic band gap ($E_g$). For each normal mode $Q_i$ in the energy region of interest, we find a linear relationship between the displacement and the band gap (i.e.\ linear electron-phonon coupling regime) and calculate the relaxation energy $\lambda_i$ as:~\cite{grisanti2013roles,coropceanu2007charge} 
\begin{equation}
\lambda_i = \left( \frac{\partial E_g}{\partial Q_i} \right)^2 \left(4\alpha_i\right)^{-1},
\label{eq:lambda}
\end{equation}
where $\alpha_i$ is the curvature of the ground state potential energy surface along the $Q_i$ normal mode of vibration. Among all normal modes below 8\,meV, our calculations identify a few with frequency similar to those showed in Fig.~\ref{fig_spec}(a) and associated with sizable relaxation energy (few meV), i.e.\ coupled to bandgap excitations. In Table~\ref{Table}, we compare the experimental vibrational frequencies with those obtained from DFT, together with the corresponding relaxation energies and a description of the characteristic atomic displacements. The agreement between calculation and measurement is remarkable, especially considering the inherent difficulty in computing normal-mode vibrations at such low frequency, where anharmonic effects can play a significant role~\cite{yaffe2017local}. The lattice motion corresponding to each of these modes is pictographically represented in Fig.~\ref{fig_spec}(b), and is also available in animated files as Supplementary Information. We have further verified the present results against calculations including spin-orbit coupling and found close agreement with the results in Table~\ref{Table} (see Table S2 in SI).   

As previously demonstrated in other lead-halide hybrid materials, the identified modes in this energy range correspond to the motion of the lead-iodide network~\cite{Leguy2016,quarti2013raman,la2015}. All the modes have contributions from rotation of the octahedra along the two pseudocubic axes lying within the inorganic sheet. With increasing energy, we also find additional contributions from the I-Pb-I bendings and I-Pb stretching or octahedral rotations orthogonal to the inorganic sheets (represented as the scissoring mode), indicative of the larger stiffness of these co-ordinates. The nature and energy of some of these modes are strikingly similar to those reported for three-dimensional perovskites based on experimental~\cite{Leguy2016,Yaffe2015,la2015} as well as theoretical~\cite{quarti2013raman} investigations.  

To further generalize the experimental findings and provide further validation for the theoretical methodology, we also investigated polycrystalline films of \ce{(NBT)2PbI4}. Fig.~\ref{fig_spec}(c) shows the integrated RISRS spectrum when the sample is photo-excited at 3.06\,eV. We observe five modes with full-widths-at-half-maximum $\leq 0.29$\,meV, distinct from the case of \ce{(PEA)2PbI4}, but again in very close agreement with the DFT predictions, see Table~\ref{Table}. All the modes appear to be shifted to higher energies, possibly due to the stiffening of the lattice induced by the octahedral distortions ubiquitous in the low-temperature phase of \ce{(NBT)2PbI4}~\cite{Cortecchia2017a}. Similar to the case of \ce{(PEA)2PbI4} we can assign the observed modes to the octahedral twist along or perpendicular to the inorganic sheet as well as the scissoring of the Pb-I-Pb angle, see Fig.~\ref{fig_spec}(d) for diagrammatic representation of the vibrations. Additional data on this material is presented in Figs.\ S8 and S9 of SI.

\begin{figure}[ht]
\centering
\includegraphics[width =0.86\textwidth]{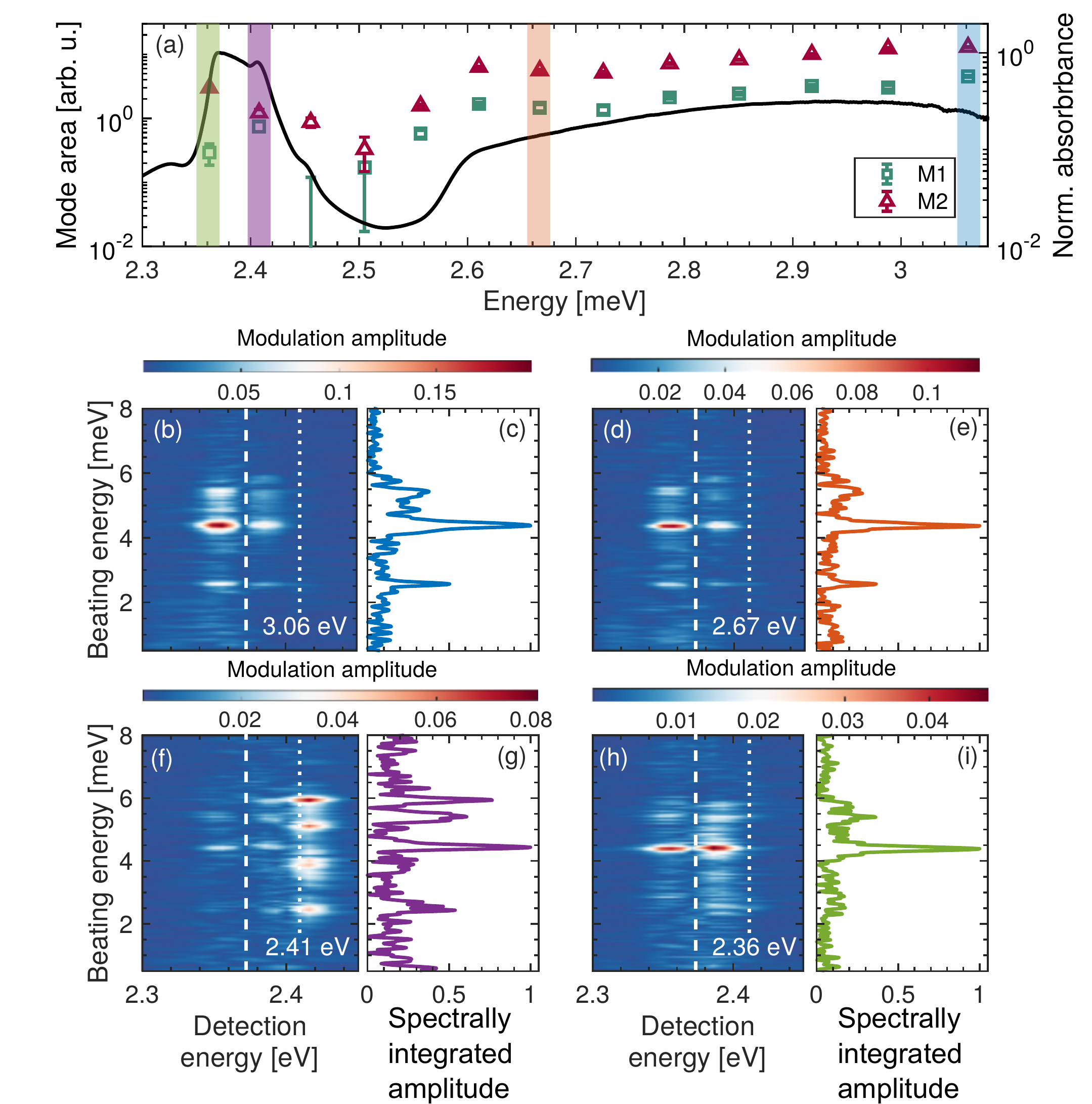}%
\caption{\textbf{Pump wavelength dependence of the resonant impulsive stimulated Raman spectra of \ce{(PEA)2PbI4} at 5\,K.} (a) Excitation profiles of modes M1 and M2 as defined in Fig.~\ref{fig_spec} (represented as symbols indicated in the inset), plotted with the linear absorption spectrum. Error bars represent statistical error arising from noise in the beating spectra. (b,d,f,h) Beating spectra as a function of detection (probe) energies. Probe-energy-integrated vibrational spectra are also shown in (c,e,g,i). The pump energies are (b/c) 3.06\,eV, (d/e) 2.67\,eV, (f/g) 2.41\,eV and (h/i) 2.36\,eV. The dashed and dotted lines over the beating maps indicate the peak energies of excitons A and B respectively, as defined in Fig.~\ref{fig1}(a).}
\label{fig2}
\end{figure}

We now return to the vibrational coherences observed in \ce{(PEA)2PbI4} to demonstrate contrasting coupling of the lattice to the two most intense excitonic transitions A and B, and also to photocarriers. Fig.~\ref{fig2}(a) shows the excitation profiles of modes M1 and M2 plotted along with the linear absorption spectrum. Due to the resonant excitation, the RISRS driving the coherent lattice motion is enhanced for those ground-state vibrational coherences that are strongly coupled to the electronic transitions\cite{dhar1994time,lanzani2008coherent}. This is clearly observed as a monotonic increase of the intensity of both the modes in the excitation spectra when the excitation energy is tuned to higher energies into the carrier continuum, at pump energies $\geq 2.56$\,eV. The photon-flux at each excitation energy is kept constant to directly correlate the mode intensity to the absorption cross-section. The full beating maps and integrated spectra corresponding to these excitation conditions are presented in Figs.\ S4 and S5 of SI. 

In Figs.~\ref{fig2} (b,d,f,h), we show the two-dimensional beating maps, which represent the probe-energy-resolved Raman spectra, obtained by Fourier transforming the dynamics at all the detection energies for each of the excitation energies marked by shaded regions in Fig.~\ref{fig2}(a). We display two spectra resulting from pumping into the continuum at 3.06 and 2.67\,eV (Figs.~\ref{fig2}(b) and \ref{fig2}(d) respectively), one spectrum resulting from pumping exciton B (Fig.~\ref{fig2}(f)), and one exciting exciton A (Fig.~\ref{fig2}(h)). Figs.~\ref{fig2} (c,e,g,i) show the integrated spectra across all the detection energies. The raw pump-probe data and cuts corresponding to these pumping conditions are presented in Figs.\ S10 to S12 of SI.

In the cases of the free-carrier excitations, we observe identical vibrational coherences with the dominant signal at modes M1 and M2. While the resonant excitation of exciton A reveals predominant coupling to mode M2, excitation of exciton B displays strikingly different vibrational spectral structure. As evident in Fig.~\ref{fig2}(g), we observe more intense signals at mode M6 and M4 along with diminished intensity of mode M2. These observations suggest rather distinct lattice couplings exhibited by each of the exciton states. Another important observation from the beating maps is the non-uniformity of the amplitude across the detection energy axis for different excitation energies. For instance, the free-carrier excitation beating maps reveal dominant signal only at the detection energies around the exciton A absorption resonance. The amplitude spectrum has a characteristic dual-peaked lineshape with a dip at the peak energy of exciton A~\cite{luer2007ultrafast}, indicated by dashed lines in the beating maps. Importantly there is no notable signal at the energy of exciton B, which is indicated by dotted lines.  The only exception to this is the case shown in Fig.~\ref{fig2}(f), where most of the signal is present around the energy of exciton B. This data corresponds to resonant excitation of exciton B, and establishes that the vibrational coherences associated with this transitions differ to those due to exciton A, as well as those generated by photocarriers. We employ this intriguing observation as the key evidence in this work to differentiate the polaronic character of each of the excitons and that of free-carrier excitations.

\begin{figure}[ht]
\centering
\includegraphics[width =0.9\textwidth]{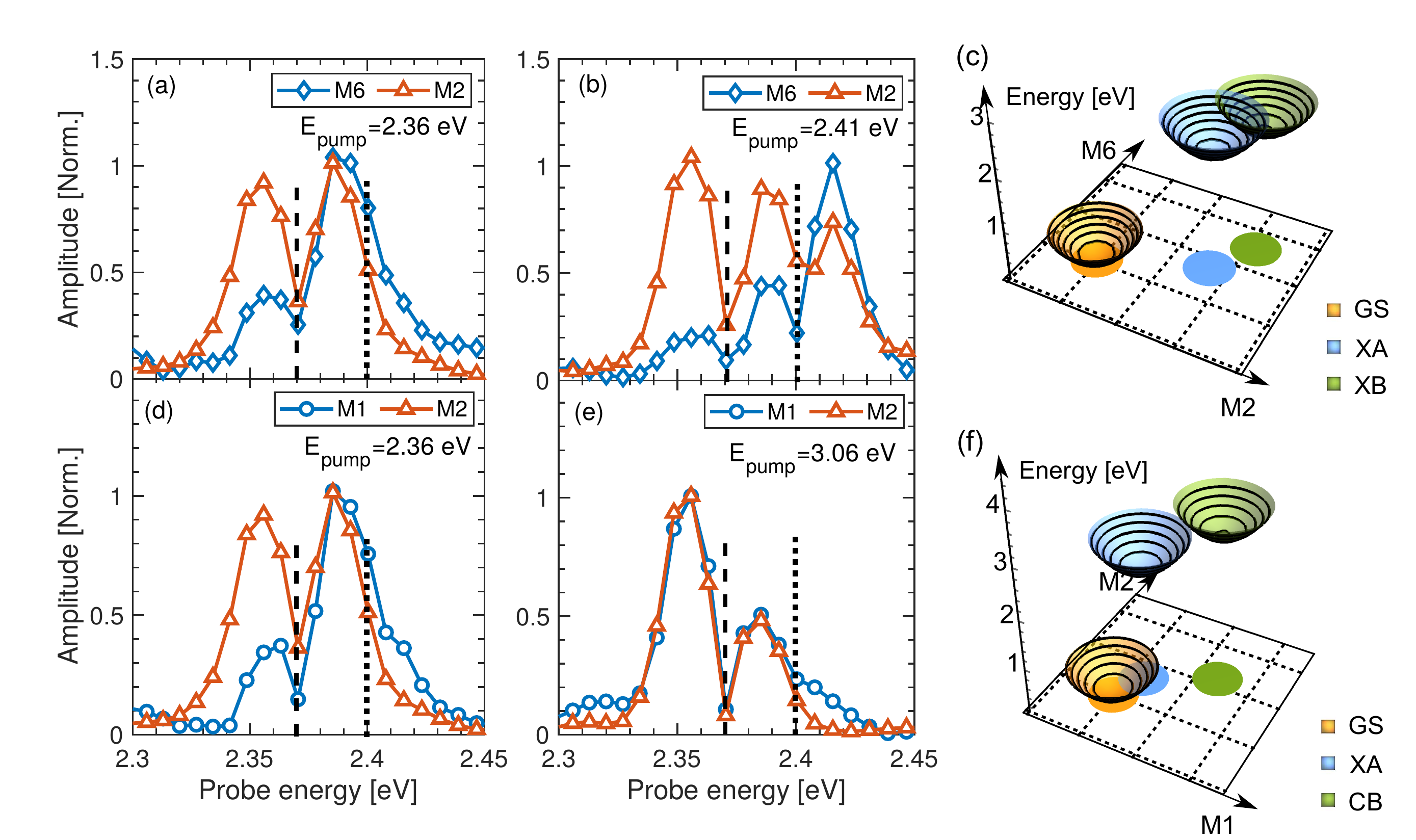}%
\caption{\textbf{Consequences of the wavepacket dynamics in \ce{(PEA)2PbI4} at 5\,K.} Amplitude spectra of M2 and M6 modes when the pump energy is (a) resonant with exciton A at 2.36\,eV  and (b) with exciton B at 2.41\,eV. (c) Configuration-space representation of the harmonic potentials of ground state (GS), exciton A and exciton B.  Amplitude spectra of M1 and M2 modes when the pump energy is (d) resonant with exciton A at 2.36\,eV and (e) with the carrier continuum at 2.41\,eV. (f) Configuration space representation of the harmonic potentials of ground state (GS), exciton A (XA, displaced predominantly along M2 coordinates) and the free carriers (CB, with strong and approximately equal displacement along M1 and M2 coordinates).}
\label{fig5}
\end{figure}

We now discuss the nature of the observed lineshapes by comparing the amplitude spectra obtained by taking horizontal cuts of the probe-energy-resolved beating maps from different cases in Fig.~\ref{fig5}. We first consider exciton A in Fig.~\ref{fig5}(a) and exciton B in Fig.~\ref{fig5}(b). Given that M2 and M6 are the dominant vibrational modes in these two cases, we plot the amplitude spectra taken at those energies. The lineshapes in Fig.~\ref{fig5}(a) are indicative of vibrational wavepacket dynamics often observed in molecular systems~\cite{lanzani2008coherent, luer2007ultrafast}, where the energy of exciton A is modulated by the motion of a vibrational wavepacket along the a real-space vibrational coordinate associated with the coherently excited phonon. The dual-peaked lineshape and the $\pi$ phase shift at the peak energy (not shown here but displayed in section S4 of the Supplemental Information) is clearly indicative of such dynamics, where the detection energies around the exciton peak display the most prominent oscillations. The observed lineshape can be reproduced using a simple harmonic oscillator model involving two electronic states, which are defined by their respective potential energy surfaces (PES). The PES minimum of the excited state is displaced along one of the vibrational normal coordinates due to electron-phonon coupling. Impulsive optical excitation generates a vibrational wavepacket, which oscillates along the PESs. The effect of these vibrational degrees of freedom on the permittivity can be computed as described in ref.~\citenum{kumar2001}. A detailed description of this  model and the results of such simulations are given in the Supplemental Information (Section S3). The observed dynamics are a consequence of the displacement of the PES of the excited state perceived via spectator lattice modes and thus are indicative of polaronic effects~\cite{Batignani2017,park2018}.  We can compare the relative displacement ($\Delta$) of the PES in the excited state across each of the normal mode coordinates. The Raman cross section is proportional to $\Delta^2\omega^2$, with $\hbar \omega$ being the phonon energy, and thus the analysis of relative intensities of each of the modes at different pump energies provides a representation of the complex excited-state landscape in \ce{(PEA)2PbI4}, as depicted schematically in Fig.~\ref{fig5}(c). By pumping exciton A, we can infer that its PES is displaced strongly along the normal coordinate defined by the phonon at M2, with a relatively more limited displacement along all the other co-ordinates, as represented by the \textit{blue} PES shown in Fig.~\ref{fig5}(c). 
The absence of any modulation at the energy of exciton B in this case suggests the relative immunity of exciton B to the wavepacket dynamics only along the co-ordinate defined by the normal mode M2. This is duly supported by the amplitude spectra shown  in Fig.~\ref{fig5}(b), where the M6 mode also modulates the exciton B transition strongly. We may infer that the PES of exciton B is displaced along both M2 and M6 axes as represented by the \textit{green} PES in Fig.~\ref{fig5}(c). 

We highlight that the PES associated with exciton A is also displaced, albeit slightly, along the coordinate axis associated with M1, as evident from the amplitude spectra shown in Fig.~\ref{fig5}(d) and represented schematically in Fig.~\ref{fig5}(f). The PES of photocarriers (Fig.~\ref{fig5}(f)), on the other hand, shows much more significant displacement along both M1 as well as M2, which can also be observed in the relatively higher intensity for these modes when the pump is tuned to the carrier continuum (see Fig.~\ref{fig2}). This suggests that the photocarrier induces substantially larger lattice reorganization than the two excitons. This can be rationalized by considering the ionic nature of the perovskite lattice, which is subjected to stronger Coulomb potential in the presence of charged photoexcitations~\cite{Emin2013}, in contrast to a globally neutral excitonic quasi-particle.    

An important question is whether the modes reported in in Table~\ref{Table} are primarily sourced by the lattice reorganization around the hole or the electron. According to Eq.~\ref{eq:lambda}, the relaxation energy is calculated from the change in the single-particle bandgap ($E_g$) when the atomic positions are displaced along the coordinates of the various normal modes $Q_i$. Using the same protocol, we can track the variations in the bandgap that are due to a shift in the valence or in the conduction band edge to assess hole and electron polaron relaxation energies, respectively. The obtained dimensionless electron-phonon couplings for holes and electrons in \ce{(PEA)2PbI4} and \ce{(NBT)2PbI4}, listed on Table S1 in SI, clearly demonstrate that for \ce{(PEA)2PbI4}, M1$^\prime$, M2, M5 and M6, contribute to the formation of hole-polarons, while M1 and M3 contribute to the formation of electron-polarons. On the other hand, for \ce{(NBT)2PbI4}, all vibrational modes are more strongly coupled to holes than electrons. The fact that changes in the electronic structure induced by reorganization of the lattice are more pronounced for the valence band edge is not surprising. The valence band is indeed primarily composed by antibonding combination of 5p atomic orbitals from iodine and 6s orbitals from lead. Thus, any change in the relative distance and orientation of the lead and iodine ions, as induced by displacement along the inorganic lattice vibrational modes, is expected to significantly affect the wavefunction overlap and Pb-I hybridization, hence energy, of the valence band. By comparison, the conduction-band edge is instead mostly composed by 6p orbitals of lead. It does, therefore, not come as a surprise that mode M3 in \ce{(PEA)2PbI4} couples more strongly to the conduction band, as this mode is associated with the displacement of Pb ions within the \ce{PbI6} octahedron, and that a change along this coordinate should strongly affect the atomic overlap among the 6p orbitals of lead. 
 
Our analysis highlights the complex landscape encountered by a charge carriers before they relax to the excitonic states, scanning across different lattice configurations. We have already demonstrated that the exciton-phonon coupling can be tuned by the nature of the organic cation and the thickness of the quantum-well~\cite{Neutzner2018,gong2018electron}. In the context of growing number of custom-designed organic molecules that are being developed to template the 2D hybrid perovskite for optoelectronics, our observation garners fundamental importance to establish the optimum relaxation pathway following charge-carrier injection. We consider that the type of quantitative development of the relevant PESs is a challenging but fundamentally important task for large-scale molecular dynamics simulations that capture accurately the nonadiabatic quantum dynamics implied by this work~\cite{neukirch2016polaron,park2018}. 

Lastly, we have unambiguously demonstrated the presence of multiple distinct excitonic transitions separated by 35\,meV~\cite{Neutzner2018}. Our observation of vastly different spectral structures in the resonant vibrational excitation spectra effectively rules out their previous assignments to vibronic progressions~\cite{Straus2016a}.  It also establishes substantial and importantly distinct polaronic character of each of the excitons. We consider that this observation strongly suggests the role of polaronic binding as at least part of the origin of the spectral fine structure, which we have considered in ref.~\citenum{Neutzner2018}. This was initially motivated by a simple estimate of the polaron binding energy that was consistent with the energy splitting of 35\,meV within the fine structure.  We highlight that the origin of the fine structure can only be rigorously established via a detailed theoretical treatment that would predict the full excitonic dispersion and that includes spin-orbit coupling effects~\cite{Zhai2017}, exchange interactions~\cite{Ema2006,Kitazawa2010a, takagi2013influence}, many-body correlations~\cite{Thouin2017} and based on our current observation, non-negligible yet complex polaronic effects~\cite{zheng1998polaronic}.  

It has been suggested in the case of bulk lead-halide perovskites that polaronic effects shield the photo-generated carriers from lossy scattering pathways involving defects, LO phonons or Auger-like processes~\cite{zhu2016screening,miyata2017large}. We identified a diminished polaronic character of 
excitons that can potentially enhance the multi-particle scattering processes. Intriguingly, in our earlier work on the multi-dimensional spectroscopy of \ce{(PEA)2PbI4}, we observed signatures of bound biexcitons with different binding energies for AA and BB, with additional evidence that A-excitons also experience repulsive interactions~\cite{Thouin2017}. The consequences of the nature of excitonic structure on these many-body physics are clearly critical in the context of, for example, biexciton lasing~\cite{Kondo1998,Booker2018}.

\section*{References}

\begin{thebibliography}{10}
\expandafter\ifx\csname url\endcsname\relax
  \def\url#1{\texttt{#1}}\fi
\expandafter\ifx\csname urlprefix\endcsname\relax\def\urlprefix{URL }\fi
\providecommand{\bibinfo}[2]{#2}
\providecommand{\eprint}[2][]{\url{#2}}

\bibitem{Ishihara1989}
\bibinfo{author}{Ishihara, T.}, \bibinfo{author}{Takahashi, J.} \&
  \bibinfo{author}{Goto, T.}
\newblock \bibinfo{title}{{Exciton state in two-dimensional perovskite
  semiconductor (C$_{10}$H$_{21}$NH$_3$)$_2$PbI$_4$}}.
\newblock \emph{\bibinfo{journal}{Solid State Commun.}}
  \textbf{\bibinfo{volume}{69}}, \bibinfo{pages}{933--936}
  (\bibinfo{year}{1989}).

\bibitem{Even2014}
\bibinfo{author}{Even, J.}, \bibinfo{author}{Pedesseau, L.} \&
  \bibinfo{author}{Katan, C.}
\newblock \bibinfo{title}{{Understanding quantum confinement of charge carriers
  in layered 2D hybrid perovskites}}.
\newblock \emph{\bibinfo{journal}{ChemPhysChem}} \textbf{\bibinfo{volume}{15}},
  \bibinfo{pages}{3733--3741} (\bibinfo{year}{2014}).

\bibitem{Even2015}
\bibinfo{author}{Even, J.} \emph{et~al.}
\newblock \bibinfo{title}{{Solid-State Physics Perspective on Hybrid Perovskite
  Semiconductors}}.
\newblock \emph{\bibinfo{journal}{J. Phys. Chem. C}}
  \textbf{\bibinfo{volume}{119}}, \bibinfo{pages}{10161--10177}
  (\bibinfo{year}{2015}).

\bibitem{Saparov2016}
\bibinfo{author}{Saparov, B.} \& \bibinfo{author}{Mitzi, D.~B.}
\newblock \bibinfo{title}{{Organic-Inorganic Perovskites: Structural
  Versatility for Functional Materials Design}}.
\newblock \emph{\bibinfo{journal}{Chem. Rev.}} \textbf{\bibinfo{volume}{116}},
  \bibinfo{pages}{4558--4596} (\bibinfo{year}{2016}).

\bibitem{blancon2018}
\bibinfo{author}{Blancon, J.-C.} \emph{et~al.}
\newblock \bibinfo{title}{Scaling law for excitons in 2d perovskite quantum
  wells}.
\newblock \emph{\bibinfo{journal}{Nature Commun.}}
  \textbf{\bibinfo{volume}{9}}, \bibinfo{pages}{2254} (\bibinfo{year}{2018}).

\bibitem{Neutzner2018}
\bibinfo{author}{Neutzner, S.} \emph{et~al.}
\newblock \bibinfo{title}{Exciton-polaron spectral structures in
  two-dimensional hybrid lead-halide perovskites}.
\newblock \emph{\bibinfo{journal}{Phys. Rev. Materials}}
  \textbf{\bibinfo{volume}{2}}, \bibinfo{pages}{064605} (\bibinfo{year}{2018}).

\bibitem{Yaffe2015}
\bibinfo{author}{Yaffe, O.} \emph{et~al.}
\newblock \bibinfo{title}{{Excitons in ultrathin organic-inorganic perovskite
  crystals}}.
\newblock \emph{\bibinfo{journal}{Phys. Rev. B}} \textbf{\bibinfo{volume}{92}},
  \bibinfo{pages}{045414} (\bibinfo{year}{2015}).

\bibitem{kandadad2016photophysics}
\bibinfo{author}{Srimath~Kandada, A.~R.} \& \bibinfo{author}{Petrozza, A.}
\newblock \bibinfo{title}{Photophysics of hybrid lead halide perovskites: The
  role of microstructure}.
\newblock \emph{\bibinfo{journal}{Acc. Chem. Res.}}
  \textbf{\bibinfo{volume}{49}}, \bibinfo{pages}{536--544}
  (\bibinfo{year}{2016}).

\bibitem{miyata2017large}
\bibinfo{author}{Miyata, K.} \emph{et~al.}
\newblock \bibinfo{title}{Large polarons in lead halide perovskites}.
\newblock \emph{\bibinfo{journal}{Sci. Adv.}} \textbf{\bibinfo{volume}{3}},
  \bibinfo{pages}{e1701217} (\bibinfo{year}{2017}).

\bibitem{straus2018_2}
\bibinfo{author}{Straus, D.~B.} \& \bibinfo{author}{Kagan, C.~R.}
\newblock \bibinfo{title}{Electrons, excitons, and phonons in two-dimensional
  hybrid perovskites: Connecting structural, optical, and electronic
  properties}.
\newblock \emph{\bibinfo{journal}{J. Phys. Chem. Lett.}}
  \textbf{\bibinfo{volume}{9}}, \bibinfo{pages}{1434--1447}
  (\bibinfo{year}{2018}).

\bibitem{Thouin2017}
\bibinfo{author}{Thouin, F.} \emph{et~al.}
\newblock \bibinfo{title}{{Stable biexcitons in two-dimensional metal-halide
  perovskites with strong dynamic lattice disorder}}.
\newblock \emph{\bibinfo{journal}{Phys. Rev. Materials}}
  \textbf{\bibinfo{volume}{2}}, \bibinfo{pages}{034001} (\bibinfo{year}{2018}).

\bibitem{Kondo1998}
\bibinfo{author}{Kondo, T.}, \bibinfo{author}{Azuma, T.},
  \bibinfo{author}{Yuasa, T.} \& \bibinfo{author}{Ito, R.}
\newblock \bibinfo{title}{{Biexciton lasing in the layered perovskite-type
  material (C$_6$H$_{13}$NH$_3$)$_2$PbI$_4$}}.
\newblock \emph{\bibinfo{journal}{Solid State Commun.}}
  \textbf{\bibinfo{volume}{105}}, \bibinfo{pages}{253--255}
  (\bibinfo{year}{1998}).

\bibitem{Quan:2017aa}
\bibinfo{author}{Quan, L.~N.} \emph{et~al.}
\newblock \bibinfo{title}{Tailoring the energy landscape in quasi-2d halide
  perovskites enables efficient green-light emission}.
\newblock \emph{\bibinfo{journal}{Nano Lett.}} \textbf{\bibinfo{volume}{17}},
  \bibinfo{pages}{3701--3709} (\bibinfo{year}{2017}).

\bibitem{Su:2017aa}
\bibinfo{author}{Su, R.} \emph{et~al.}
\newblock \bibinfo{title}{{Room-Temperature Polariton Lasing in All-Inorganic
  Perovskite Nanoplatelets}}.
\newblock \emph{\bibinfo{journal}{Nano Lett.}} \textbf{\bibinfo{volume}{17}},
  \bibinfo{pages}{3982--3988} (\bibinfo{year}{2017}).

\bibitem{Booker2018}
\bibinfo{author}{Booker, E.~P.} \emph{et~al.}
\newblock \bibinfo{title}{Vertical cavity biexciton lasing in {2D}
  dodecylammonium lead iodide perovskites}.
\newblock \emph{\bibinfo{journal}{Adv. Opt. Mater.}} \bibinfo{pages}{1800616}
  (\bibinfo{year}{2018}).

\bibitem{Senger2003aa}
\bibinfo{author}{Senger, R.~T.} \& \bibinfo{author}{Bajaj, K.~K.}
\newblock \bibinfo{title}{Binding energies of excitons in polar quantum well
  heterostructures}.
\newblock \emph{\bibinfo{journal}{Phys. Rev. B}} \textbf{\bibinfo{volume}{68}},
  \bibinfo{pages}{205314} (\bibinfo{year}{2003}).

\bibitem{Dvorak2013aa}
\bibinfo{author}{Dvorak, M.}, \bibinfo{author}{Wei, S.-H.} \&
  \bibinfo{author}{Wu, Z.}
\newblock \bibinfo{title}{Origin of the variation of exciton binding energy in
  semiconductors}.
\newblock \emph{\bibinfo{journal}{Phys. Rev. Lett.}}
  \textbf{\bibinfo{volume}{110}}, \bibinfo{pages}{016402}
  (\bibinfo{year}{2013}).

\bibitem{Ishihara1990}
\bibinfo{author}{Ishihara, T.}, \bibinfo{author}{Takahashi, J.} \&
  \bibinfo{author}{Goto, T.}
\newblock \bibinfo{title}{{Optical properties due to electronic transitions in
  two-dimensional semiconductors (C$_n$H$_{2n+1}$NH$_3$)$_2$PbI$_4$}}.
\newblock \emph{\bibinfo{journal}{Phys. Rev. B}} \textbf{\bibinfo{volume}{42}},
  \bibinfo{pages}{11099--11107} (\bibinfo{year}{1990}).

\bibitem{kataoka1993magneto}
\bibinfo{author}{Kataoka, T.} \emph{et~al.}
\newblock \bibinfo{title}{{Magneto-optical study on excitonic spectra in
  (C$_6$H$_{13}$NH$_3$)$_2$PbI$_4$}}.
\newblock \emph{\bibinfo{journal}{Phys. Rev. B}} \textbf{\bibinfo{volume}{47}},
  \bibinfo{pages}{2010--2018} (\bibinfo{year}{1993}).

\bibitem{Tanaka2005}
\bibinfo{author}{Tanaka, K.} \emph{et~al.}
\newblock \bibinfo{title}{{Electronic and excitonic structures of
  inorganic-organic perovskite-type quantum-well crystal
  (C$_4$H$_9$NH$_3$)$_2$PbBr$_4$}}.
\newblock \emph{\bibinfo{journal}{Jap. J. Appl. Phys., Part 1}}
  \textbf{\bibinfo{volume}{44}}, \bibinfo{pages}{5923--5932}
  (\bibinfo{year}{2005}).

\bibitem{Shimizu2005}
\bibinfo{author}{Shimizu, M.}, \bibinfo{author}{Fujisawa, J.~I.} \&
  \bibinfo{author}{Ishi-Hayase, J.}
\newblock \bibinfo{title}{{Influence of dielectric confinement on excitonic
  nonlinearity in inorganic-organic layered semiconductors}}.
\newblock \emph{\bibinfo{journal}{Phys. Rev. B}} \textbf{\bibinfo{volume}{71}},
  \bibinfo{pages}{1--9} (\bibinfo{year}{2005}).

\bibitem{Ema2006}
\bibinfo{author}{Ema, K.} \emph{et~al.}
\newblock \bibinfo{title}{{Huge exchange energy and fine structure of excitons
  in an organic-inorganic quantum well material}}.
\newblock \emph{\bibinfo{journal}{Phys. Rev. B}} \textbf{\bibinfo{volume}{73}},
  \bibinfo{pages}{3--6} (\bibinfo{year}{2006}).

\bibitem{Goto2006}
\bibinfo{author}{Goto, T.} \emph{et~al.}
\newblock \bibinfo{title}{{Localization of triplet excitons and biexcitons in
  the two-dimensional semiconductor
  (CH$_3$C$_6$H$_4$CH$_2$NH$_3$)$_2$PbBr$_4$}}.
\newblock \emph{\bibinfo{journal}{Phys. Rev. B}} \textbf{\bibinfo{volume}{73}},
  \bibinfo{pages}{115206} (\bibinfo{year}{2006}).

\bibitem{Kitazawa2010a}
\bibinfo{author}{Kitazawa, N.} \& \bibinfo{author}{Watanabe, Y.}
\newblock \bibinfo{title}{{Optical properties of natural quantum-well compounds
  (C$_6$H$_5$-C$_n$H$_{2n}$-NH$_3$)$_2$PbBr$_4$ ($n$=1--4)}}.
\newblock \emph{\bibinfo{journal}{J. Phys. Chem. Solids}}
  \textbf{\bibinfo{volume}{71}}, \bibinfo{pages}{797--802}
  (\bibinfo{year}{2010}).

\bibitem{Gauthron:10}
\bibinfo{author}{Gauthron, K.} \emph{et~al.}
\newblock \bibinfo{title}{{Optical spectroscopy of two-dimensional layered
  (C$_6$H$_5$C$_2$H$_4$-NH$_3$)$_2$-PbI$_4$ perovskite}}.
\newblock \emph{\bibinfo{journal}{Opt. Express}} \textbf{\bibinfo{volume}{18}},
  \bibinfo{pages}{5912--5919} (\bibinfo{year}{2010}).

\bibitem{Straus2016a}
\bibinfo{author}{Straus, D.~B.} \emph{et~al.}
\newblock \bibinfo{title}{{Direct Observation of Electron-Phonon Coupling and
  Slow Vibrational Relaxation in Organic-Inorganic Hybrid Perovskites}}.
\newblock \emph{\bibinfo{journal}{J. Am. Chem. Soc.}}
  \textbf{\bibinfo{volume}{138}}, \bibinfo{pages}{13798--13801}
  (\bibinfo{year}{2016}).

\bibitem{quarti2018tuning}
\bibinfo{author}{Quarti, C.}, \bibinfo{author}{Marchal, N.} \&
  \bibinfo{author}{Beljonne, D.}
\newblock \bibinfo{title}{Tuning the optoelectronic properties of 2d hybrid
  perovskite semiconductors with alkyl chain spacers}.
\newblock \emph{\bibinfo{journal}{J. Phys. Chem. Lett}}
  \textbf{\bibinfo{volume}{9}}, \bibinfo{pages}{3416 -- 3424}
  (\bibinfo{year}{2018}).

\bibitem{sood1985resonance}
\bibinfo{author}{Sood, A.}, \bibinfo{author}{Menendez, J.},
  \bibinfo{author}{Cardona, M.} \& \bibinfo{author}{Ploog, K.}
\newblock \bibinfo{title}{{Resonance Raman scattering by confined LO and TO
  phonons in GaAs-AlAs superlattices}}.
\newblock \emph{\bibinfo{journal}{Phys. Rev. Lett.}}
  \textbf{\bibinfo{volume}{54}}, \bibinfo{pages}{2111} (\bibinfo{year}{1985}).

\bibitem{dhar1994time}
\bibinfo{author}{Dhar, L.}, \bibinfo{author}{Rogers, J.~A.} \&
  \bibinfo{author}{Nelson, K.~A.}
\newblock \bibinfo{title}{Time-resolved vibrational spectroscopy in the
  impulsive limit}.
\newblock \emph{\bibinfo{journal}{Chem. Rev.}} \textbf{\bibinfo{volume}{94}},
  \bibinfo{pages}{157--193} (\bibinfo{year}{1994}).

\bibitem{merlin1997generating}
\bibinfo{author}{Merlin, R.}
\newblock \bibinfo{title}{Generating coherent thz phonons with light pulses}.
\newblock \emph{\bibinfo{journal}{Solid State Commun.}}
  \textbf{\bibinfo{volume}{102}}, \bibinfo{pages}{207--220}
  (\bibinfo{year}{1997}).

\bibitem{Cortecchia2017a}
\bibinfo{author}{Cortecchia, D.} \emph{et~al.}
\newblock \bibinfo{title}{{Broadband Emission in Two-Dimensional Hybrid
  Perovskites: The Role of Structural Deformation}}.
\newblock \emph{\bibinfo{journal}{J. Am. Chem. Soc.}}
  \textbf{\bibinfo{volume}{139}}, \bibinfo{pages}{39--42}
  (\bibinfo{year}{2017}).

\bibitem{guo2016electron}
\bibinfo{author}{Guo, Z.}, \bibinfo{author}{Wu, X.}, \bibinfo{author}{Zhu, T.},
  \bibinfo{author}{Zhu, X.} \& \bibinfo{author}{Huang, L.}
\newblock \bibinfo{title}{Electron--phonon scattering in atomically thin 2d
  perovskites}.
\newblock \emph{\bibinfo{journal}{ACS nano}} \textbf{\bibinfo{volume}{10}},
  \bibinfo{pages}{9992--9998} (\bibinfo{year}{2016}).

\bibitem{grancini2015role}
\bibinfo{author}{Grancini, G.} \emph{et~al.}
\newblock \bibinfo{title}{Role of microstructure in the electron--hole
  interaction of hybrid lead halide perovskites}.
\newblock \emph{\bibinfo{journal}{Nat. Photon.}} \textbf{\bibinfo{volume}{9}},
  \bibinfo{pages}{695} (\bibinfo{year}{2015}).

\bibitem{Haug2008}
\bibinfo{author}{Haug, H.} \& \bibinfo{author}{Koch, S.~W.}
\newblock \emph{\bibinfo{title}{{Quantum theory of the optical and electronic
  properties of semiconductors}}} (\bibinfo{publisher}{World Scientific},
  \bibinfo{address}{London}, \bibinfo{year}{2008}).

\bibitem{ivanovska2016vibrational}
\bibinfo{author}{Ivanovska, T.} \emph{et~al.}
\newblock \bibinfo{title}{Vibrational response of methylammonium lead iodide:
  From cation dynamics to phonon--phonon interactions}.
\newblock \emph{\bibinfo{journal}{ChemSusChem}} \textbf{\bibinfo{volume}{9}},
  \bibinfo{pages}{2994--3004} (\bibinfo{year}{2016}).

\bibitem{corno2006periodic}
\bibinfo{author}{Corno, M.}, \bibinfo{author}{Busco, C.},
  \bibinfo{author}{Civalleri, B.} \& \bibinfo{author}{Ugliengo, P.}
\newblock \bibinfo{title}{Periodic ab initio study of structural and
  vibrational features of hexagonal hydroxyapatite
  ca$_10$(po$_4$)$_6$(oh)$_2$}.
\newblock \emph{\bibinfo{journal}{Phys. Chem. Chem. Phys.}}
  \textbf{\bibinfo{volume}{8}}, \bibinfo{pages}{2464--2472}
  (\bibinfo{year}{2006}).

\bibitem{brivio2015lattice}
\bibinfo{author}{Brivio, F.} \emph{et~al.}
\newblock \bibinfo{title}{Lattice dynamics and vibrational spectra of the
  orthorhombic, tetragonal, and cubic phases of methylammonium lead iodide}.
\newblock \emph{\bibinfo{journal}{Phys. Rev. B}} \textbf{\bibinfo{volume}{92}},
  \bibinfo{pages}{144308} (\bibinfo{year}{2015}).

\bibitem{quarti2013raman}
\bibinfo{author}{Quarti, C.} \emph{et~al.}
\newblock \bibinfo{title}{{The Raman spectrum of the CH$_3$NH$_3$PbI$_3$ hybrid
  perovskite: interplay of theory and experiment}}.
\newblock \emph{\bibinfo{journal}{J. Phys. Chem. Lett.}}
  \textbf{\bibinfo{volume}{5}}, \bibinfo{pages}{279--284}
  (\bibinfo{year}{2013}).

\bibitem{grisanti2013roles}
\bibinfo{author}{Grisanti, L.} \emph{et~al.}
\newblock \bibinfo{title}{Roles of local and nonlocal electron-phonon couplings
  in triplet exciton diffusion in the anthracene crystal}.
\newblock \emph{\bibinfo{journal}{Phys. Rev. B}} \textbf{\bibinfo{volume}{88}},
  \bibinfo{pages}{035450} (\bibinfo{year}{2013}).

\bibitem{coropceanu2007charge}
\bibinfo{author}{Coropceanu, V.} \emph{et~al.}
\newblock \bibinfo{title}{Charge transport in organic semiconductors}.
\newblock \emph{\bibinfo{journal}{Chem. Rev.}} \textbf{\bibinfo{volume}{107}},
  \bibinfo{pages}{926--952} (\bibinfo{year}{2007}).

\bibitem{yaffe2017local}
\bibinfo{author}{Yaffe, O.} \emph{et~al.}
\newblock \bibinfo{title}{Local polar fluctuations in lead halide perovskite
  crystals}.
\newblock \emph{\bibinfo{journal}{Phys. Rev. Lett.}}
  \textbf{\bibinfo{volume}{118}}, \bibinfo{pages}{136001}
  (\bibinfo{year}{2017}).

\bibitem{Leguy2016}
\bibinfo{author}{Leguy, A. M.~A.} \emph{et~al.}
\newblock \bibinfo{title}{{Dynamic disorder, phonon lifetimes, and the
  assignment of modes to the vibrational spectra of methylammonium lead halide
  perovskites}}.
\newblock \emph{\bibinfo{journal}{Phys. Chem. Chem. Phys.}}
  \textbf{\bibinfo{volume}{18}}, \bibinfo{pages}{27051--27066}
  (\bibinfo{year}{2016}).

\bibitem{la2015}
\bibinfo{author}{La-O-Vorakiat, C.} \emph{et~al.}
\newblock \bibinfo{title}{Phonon mode transformation across the
  orthohombic--tetragonal phase transition in a lead iodide perovskite
  {CH$_3$NH$_3$PbI$_3$}: A terahertz time-domain spectroscopy approach}.
\newblock \emph{\bibinfo{journal}{J. Phys. Chem. Lett.}}
  \textbf{\bibinfo{volume}{7}}, \bibinfo{pages}{1--6} (\bibinfo{year}{2015}).

\bibitem{lanzani2008coherent}
\bibinfo{author}{De~Silvestri, S.}, \bibinfo{author}{Cerullo, G.} \&
  \bibinfo{author}{Lanzani, G.}
\newblock \emph{\bibinfo{title}{Coherent vibrational dynamics}}
  (\bibinfo{publisher}{CRC Press}, \bibinfo{year}{2008}).

\bibitem{luer2007ultrafast}
\bibinfo{author}{L{\"u}er, L.} \emph{et~al.}
\newblock \bibinfo{title}{Coherent phonon dynamics in semiconducting carbon
  nanotubes: A quantitative study of electron-phonon coupling}.
\newblock \emph{\bibinfo{journal}{Phys. Rev. Lett.}}
  \textbf{\bibinfo{volume}{102}}, \bibinfo{pages}{127401}
  (\bibinfo{year}{2009}).

\bibitem{kumar2001}
\bibinfo{author}{Kumar, A.~T.}, \bibinfo{author}{Rosca, F.},
  \bibinfo{author}{Widom, A.} \& \bibinfo{author}{Champion, P.~M.}
\newblock \bibinfo{title}{Investigations of amplitude and phase excitation
  profiles in femtosecond coherence spectroscopy}.
\newblock \emph{\bibinfo{journal}{J. Chem. Phys.}}
  \textbf{\bibinfo{volume}{114}}, \bibinfo{pages}{701--724}
  (\bibinfo{year}{2001}).

\bibitem{Batignani2017}
\bibinfo{author}{Batignani, G.} \emph{et~al.}
\newblock \bibinfo{title}{Probing femtosecond lattice displacement upon
  photo-carrier generation in lead halide perovskite}.
\newblock \emph{\bibinfo{journal}{Nature Commun.}}
  \textbf{\bibinfo{volume}{9}}, \bibinfo{pages}{1971} (\bibinfo{year}{2018}).

\bibitem{park2018}
\bibinfo{author}{Park, M.} \emph{et~al.}
\newblock \bibinfo{title}{Excited-state vibrational dynamics toward the polaron
  in methylammonium lead iodide perovskite}.
\newblock \emph{\bibinfo{journal}{Nature Commun.}}
  \textbf{\bibinfo{volume}{9}}, \bibinfo{pages}{2525} (\bibinfo{year}{2018}).

\bibitem{Emin2013}
\bibinfo{author}{Emin, D.}
\newblock \emph{\bibinfo{title}{{Polarons}}} (\bibinfo{publisher}{Cambridge
  University press}, \bibinfo{year}{2013}).

\bibitem{gong2018electron}
\bibinfo{author}{Gong, X.} \emph{et~al.}
\newblock \bibinfo{title}{Electron--phonon interaction in efficient perovskite
  blue emitters}.
\newblock \emph{\bibinfo{journal}{Nat. Mater.}} \textbf{\bibinfo{volume}{17}},
  \bibinfo{pages}{550--556} (\bibinfo{year}{2018}).

\bibitem{neukirch2016polaron}
\bibinfo{author}{Neukirch, A.~J.} \emph{et~al.}
\newblock \bibinfo{title}{Polaron stabilization by cooperative lattice
  distortion and cation rotations in hybrid perovskite materials}.
\newblock \emph{\bibinfo{journal}{Nano letters}} \textbf{\bibinfo{volume}{16}},
  \bibinfo{pages}{3809--3816} (\bibinfo{year}{2016}).

\bibitem{Zhai2017}
\bibinfo{author}{Zhai, Y.} \emph{et~al.}
\newblock \bibinfo{title}{{Giant Rashba splitting in 2D organic-inorganic
  halide perovskites measured by transient spectroscopies}}.
\newblock \emph{\bibinfo{journal}{Sci. Adv.}} \textbf{\bibinfo{volume}{3}},
  \bibinfo{pages}{e1700704} (\bibinfo{year}{2017}).

\bibitem{takagi2013influence}
\bibinfo{author}{Takagi, H.}, \bibinfo{author}{Kunugita, H.} \&
  \bibinfo{author}{Ema, K.}
\newblock \bibinfo{title}{Influence of the image charge effect on excitonic
  energy structure in organic-inorganic multiple quantum well crystals}.
\newblock \emph{\bibinfo{journal}{Phys. Rev. B}} \textbf{\bibinfo{volume}{87}},
  \bibinfo{pages}{125421} (\bibinfo{year}{2013}).

\bibitem{zheng1998polaronic}
\bibinfo{author}{Zheng, R.} \& \bibinfo{author}{Matsuura, M.}
\newblock \bibinfo{title}{Polaronic effects on excitons in quantum wells}.
\newblock \emph{\bibinfo{journal}{Phys. Rev. B}} \textbf{\bibinfo{volume}{57}},
  \bibinfo{pages}{1749} (\bibinfo{year}{1998}).

\bibitem{zhu2016screening}
\bibinfo{author}{Zhu, H.} \emph{et~al.}
\newblock \bibinfo{title}{Screening in crystalline liquids protects energetic
  carriers in hybrid perovskites}.
\newblock \emph{\bibinfo{journal}{Science}} \textbf{\bibinfo{volume}{353}},
  \bibinfo{pages}{1409--1413} (\bibinfo{year}{2016}).

\bibitem{Calabrese1991}
\bibinfo{author}{Calabrese, J.} \emph{et~al.}
\newblock \bibinfo{title}{Preparation and characterization of layered lead
  halide compounds}.
\newblock \emph{\bibinfo{journal}{J. Am. Chem. Soc.}}
  \textbf{\bibinfo{volume}{113}}, \bibinfo{pages}{2328--2330}
  (\bibinfo{year}{1991}).

\bibitem{dovesi2018quantum}
\bibinfo{author}{Dovesi, R.} \emph{et~al.}
\newblock \bibinfo{title}{Quantum-mechanical condensed matter simulations with
  crystal}.
\newblock \emph{\bibinfo{journal}{Wiley Interdisciplinary Reviews:
  Computational Molecular Science}} \bibinfo{pages}{e1360}
  (\bibinfo{year}{2018}).

\bibitem{perdew1996generalized}
\bibinfo{author}{Perdew, J.~P.}, \bibinfo{author}{Burke, K.} \&
  \bibinfo{author}{Ernzerhof, M.}
\newblock \bibinfo{title}{Generalized gradient approximation made simple}.
\newblock \emph{\bibinfo{journal}{Phys. Rev. Lett.}}
  \textbf{\bibinfo{volume}{77}}, \bibinfo{pages}{3865} (\bibinfo{year}{1996}).

\bibitem{monkhorst1976special}
\bibinfo{author}{Monkhorst, H.~J.} \& \bibinfo{author}{Pack, J.~D.}
\newblock \bibinfo{title}{Special points for brillouin-zone integrations}.
\newblock \emph{\bibinfo{journal}{Phys. Rev. B}} \textbf{\bibinfo{volume}{13}},
  \bibinfo{pages}{5188} (\bibinfo{year}{1976}).

\end{thebibliography}

\newpage

\begin{acknowledgements}
A.R.S.K.\ acknowledges funding from EU Horizon 2020 via a Marie Sklodowska Curie Fellowship (Global) (Project No.\ 705874). F.T.\ acknowledges support from a doctoral postgraduate scholarship from the Natural Sciences and Engineering Research Council of Canada and Fond Qu\'eb\'ecois pour la Recherche: Nature et Technologies. This work is partially supported by the National Science Foundation (Award 1838276). C.S.\ acknowledges support from the School of Chemistry and Biochemistry and the College of Science of Georgia Institute of Technology. The work at Mons was supported by the Interuniversity Attraction Pole program of the Belgian Federal Science Policy Office (PAI 6/27) and FNRS-F.R.S. Computational resources have been provided by the Consortium des \'Equipements de Calcul Intensif (C\'ECI), funded by the Fonds de la Recherche Scientifique de Belgique (F.R.S.-FNRS) under Grant No.\ 2.5020.11. D.B. is a FNRS Research Director.
\end{acknowledgements}

\section*{Author Contributions}
F.T., D.A.V.C., I.B. and A.R.S.K.\ carried out transient absorption measurements. F.T.\ and D.A.V.C.\ performed the analysis of the experimental data. C.Q.\ performed ab initio calculations. D.C.\ synthesized the samples. A.P.\ supervised the sample preparation activity, D.B.\ supervised the ab initio calculations, and C.S.\ and A.R.S.K.\ supervised the ultrafast spectroscopy activity. A.R.S.K.\ and C.S.\ conceived the project. All authors contributed to the redaction of the manuscript. F.T.\ and D.A.V.C.\ are to be considered first co-authors, and C.S.\ and A.R.S.K.\ corresponding co-authors.

\section*{Competing interests}
The authors declare no competing interests.

\section*{Additional Information}

\textbf{Supplementary information} is available for this paper at [URL to be added by editor].

\textbf{Reprints and permissions information} is available at www.nature.com/reprints.

\textbf{Correspondence and requests for material} should be addressed to A.R.S.K.\ (E-mail: Srinivasa.Srimath@iit.it) or C.S. (E-mail: carlos.silva@gatech.edu).

\section*{Methods}
\subsection*{Sample preparation}
For the preparation of \ce{(PEA)2PbI4} thin films (PEA = phenylethylammonium), the precursor solution (0.25\,M) of \ce{(PEA)2PbI4} was prepared by mixing \ce{(PEA)I} (Dyesol) with \ce{PbI2} in 1:1 ratio in N,N-dimethylformamide(DMF). For example, 62.3\,mg of \ce{(PEA)I} and 57.6\,mg of \ce{PbI2} were dissolved in 500\,$\mu$L of DMF (anhydrous, Sigma Aldrich). The thin films were prepared by spin coating the precursor solutions on fused-silica substrates at 4000\,rpm, 30\,s, followed by annealing at 100$^o$C for 30\,min. The extensive structural characterization of these films are reported in our earlier works~\cite{Thouin2017,Neutzner2018}.

\subsection*{Ultrafast differential transmission measurements}
Differential Transmission spectroscopy measurements were performed using an ultrafast laser system (Pharos Model PH1-20-0200-02-10, Light Conversion) emitting 1030-nm pulses at 100\,KHz, with an output power of 20\,W and pulse duration of $\sim 220$\,fs. Experiments were carried out in an integrated transient absorption/time-resolved photoluminescence commercial setup (Light Conversion Hera). Pump wavelengths in the spectral range 360--2600\,nm (see Fig. S13 of SI for typical pump excitation spectra) were generated by feeding 10\,W from the laser output to a commercial optical parametric amplifier (Orpheus, Light Conversion, Lithuania), while 2\,W are focused onto a sapphire crystal to obtain a single-filament white-light continuum covering the spectral range $\sim 490-1050$\,nm for the probe beam.  When higher energy probe light was required, a blue white-light continuum was similarly obtained by using the second harmonic of the laser output instead. The probe beam transmitted through the sample is detected by an imaging spectrograph (Shamrock 193i, Andor Technology Ltd., UK)  in combination with a multichannel detector (256 pixels, 200--1100-nm wavelength sensitivity range). Energy densities used vary in the range 25--1100\,nJ/cm$^2$, most of the measurements were carried out at 215\,nJ/cm$^2$;  with a typical spot diameter of 1.9\,mm estimated at the 1/e$^2$ plane). Beating maps and integrated spectra corresponding to these fluences are presented in Figs. S6 and S7 of SI. All measurements were carried in a vibration-free closed-cycle cryostation (Montana Instruments). We disclose that we observed a slow degradation of the sample over long exposure to the laser light. Such a process, also widely reported for three-dimensional perovskites, appears to be reversible and can be slightly negated by photo-exposing the sample for an hour prior to the experiment. While the extent of degradation is not substantial enough to make the observed trends unreliable, the shape of the excitation spectrum should nevertheless be considered only as qualitative. Nevertheless, we consider that the comparison of excitation spectrum for different modes is rigorous.

\subsection*{Density functional theory calculations}
The present calculations are based on the harmonic approximation, which solely relies on the availability of realistic crystallographic models that, in the present case, are provided by established X-ray diffraction measurements~\cite{Calabrese1991, Cortecchia2017a, Thouin2017}. The crystalline model is relaxed using the van der Waals corrected DFT-D2 method. The Hessian matrix of the forces is then calculated on the fully relaxed structure and diagonalized to obtain the vibrational frequencies. The calculations have been performed by adopting periodic boundary conditions and localized atomic basis set as implemented in the CRYSTAL17 program~\cite{dovesi2018quantum}. The computational set-up consists of double split quality basis sets which include polarization, along with the PBE functional for the description of the exchange-correlation~\cite{perdew1996generalized}. An automatic 4x4x1 sampling of the first Brillouin zone was selected~\cite{monkhorst1976special}, where the less dense sampling is related to the direction associated to the inorganic-sheet stacking, in the reciprocal lattice. The Grimme-D2 approach was included, to improve the description of the atomic forces between the organic cations. The SCF accuracy has been increased to 10$^{10}$ Hartree, to obtain accurate interatomic forces. This computational set-up has been already tested for the parental \ce{CH3NH3PbI3} perovskite in Ref.~\citenum{ivanovska2016vibrational}  and resulted in DFT vibrational spectra in excellent agreement with the experimental data available.

\section*{Data Availability}
The experimental data and analysis material that support the findings of this study are available in the Scholarly Materials And Research @ Georgia Tech repository (SMARTech), https://smartech.gatech.edu.

\newpage
\clearpage


\section*{Supplementary material}

# Phonon coherences reveal the polaronic character of excitons in two-dimensional lead-halide perovskites

Félix Thouin,[1] David A. Valverde-Chávez,[2] Claudio Quarti,[3]
Daniele Cortecchia,[4] Ilaria Bargigia,[2] David Beljonne,[3]
Annamaria Petrozza,[4] Carlos Silva,[2,1] and Ajay Ram Srimath Kandada[1,2,4]

[1]*School of Physics, Georgia Institute of Technology,*
*837 State Street NW, Atlanta, Georgia 30332, USA*

[2]*School of Chemistry and Biochemistry, Georgia Institute of Technology,*
*901 Atlantic Drive NW, Atlanta, Georgia 30332, USA*

[3]*Laboratory for Chemistry of Novel Materials,*
*Department of Chemistry, Université de Mons,*
*Place du Parc 20, 7000, Mons, Belgium*

[4]*Center for Nano Science and Technology@PoliMi,*
*Istituto Italiano di Tecnologia, via Giovanni Pascoli 70/3, 20133 Milano, Italy*

(Dated: November 21, 2018)

## S1. X-RAY DIFFRACTION ON (PEA)$_2$PBI$_4$

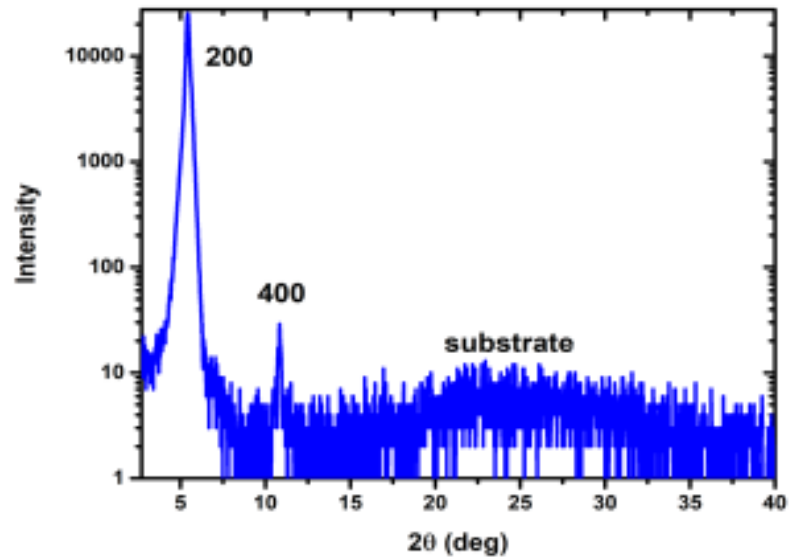

FIG. S1. X-Ray diffraction on (PEA)$_2$PbI$_4$

(PEA)$_2$PbI$_4$ films spin-coated on glass were characterized by X-Ray diffraction (XRD). The thin film XRD pattern is consistent with the structure of (PEA)$_2$PbI$_4$ (monoclinic, space group C2/m). Only the $<200>$ and $<400>$ reflections are visible in the diffractogram at 5.44° and 10.84° respectively, indicating a strong preferential orientation of the perovskite towards the [h00] direction (perovskites planes parallel to the substrate). The absence of diffraction peaks of secondary phases indicates the excellent purity of the material.

## S2. COMPLETE DATASETS

### A. Full temperature dependent dataset on $(PEA)_2PbI_4$

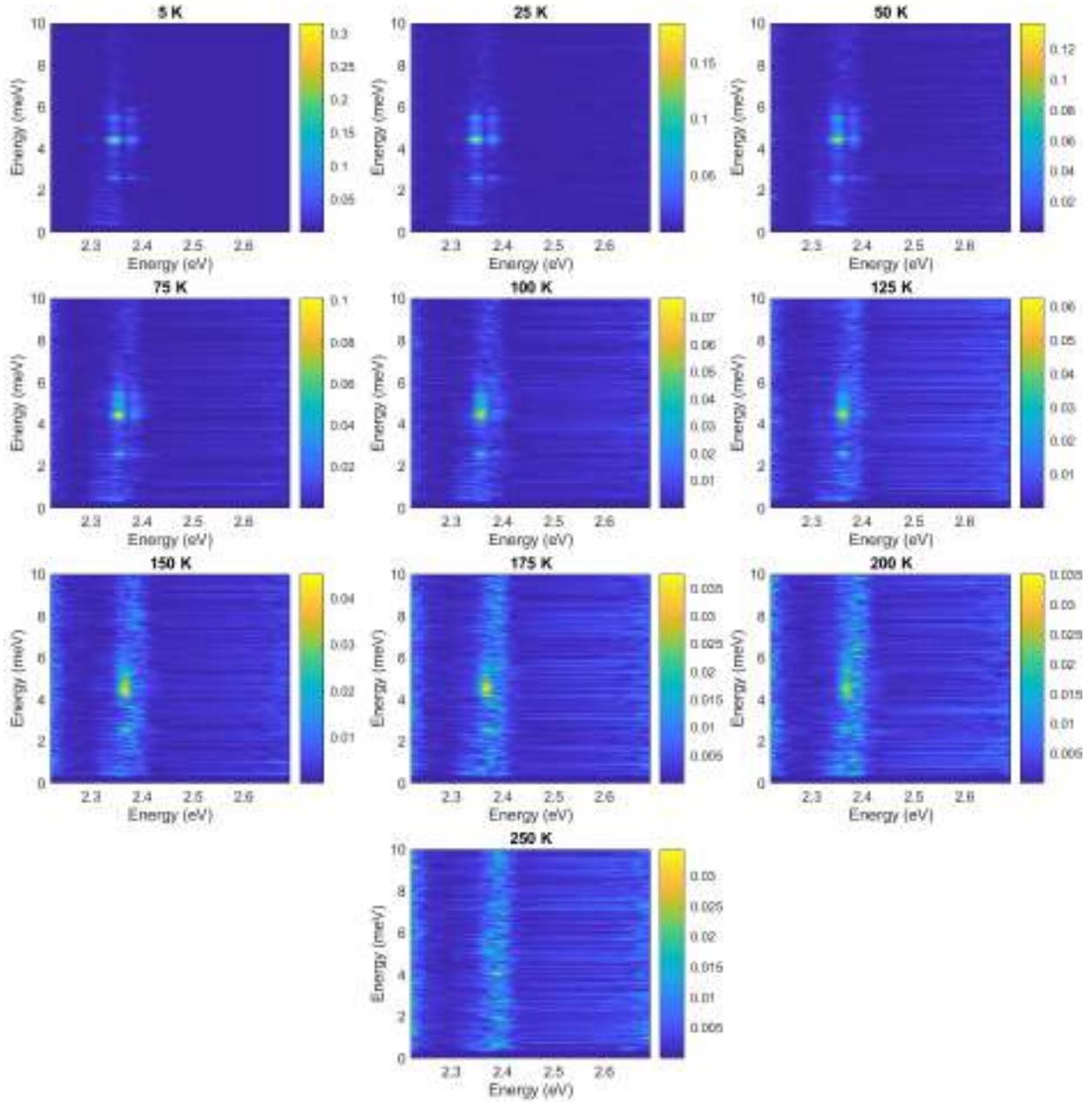

FIG. S2. Temperature dependent beating maps dataset from 5K to 250K when pumping at 3.06 eV. The sample's temperature is indicated in the title of each subfigure.

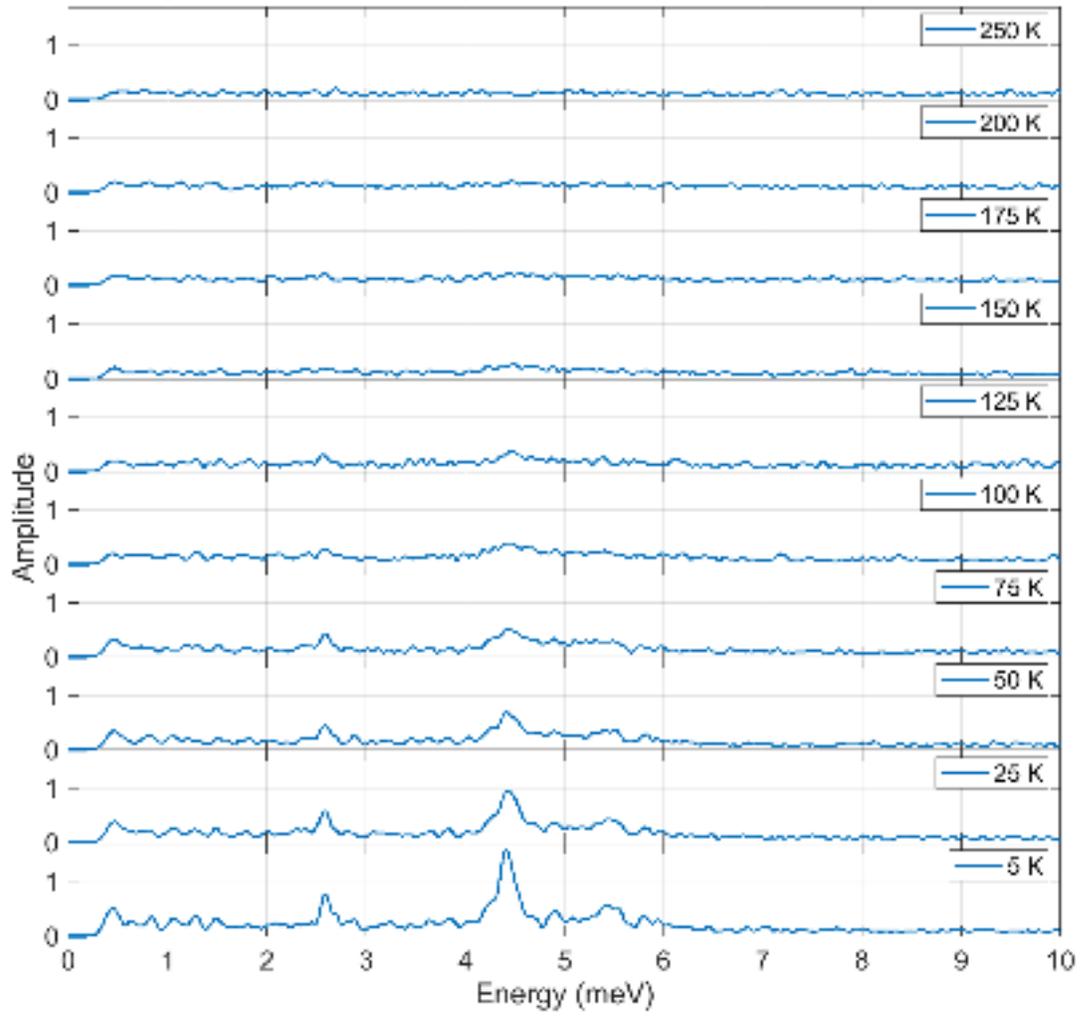

FIG. S3. Temperature dependence dataset from 5 K to 250 K, integrated along the probe axis.

## B. Full pump-wavelength dependence beating on $(PEA)_2PbI_4$ at 5K

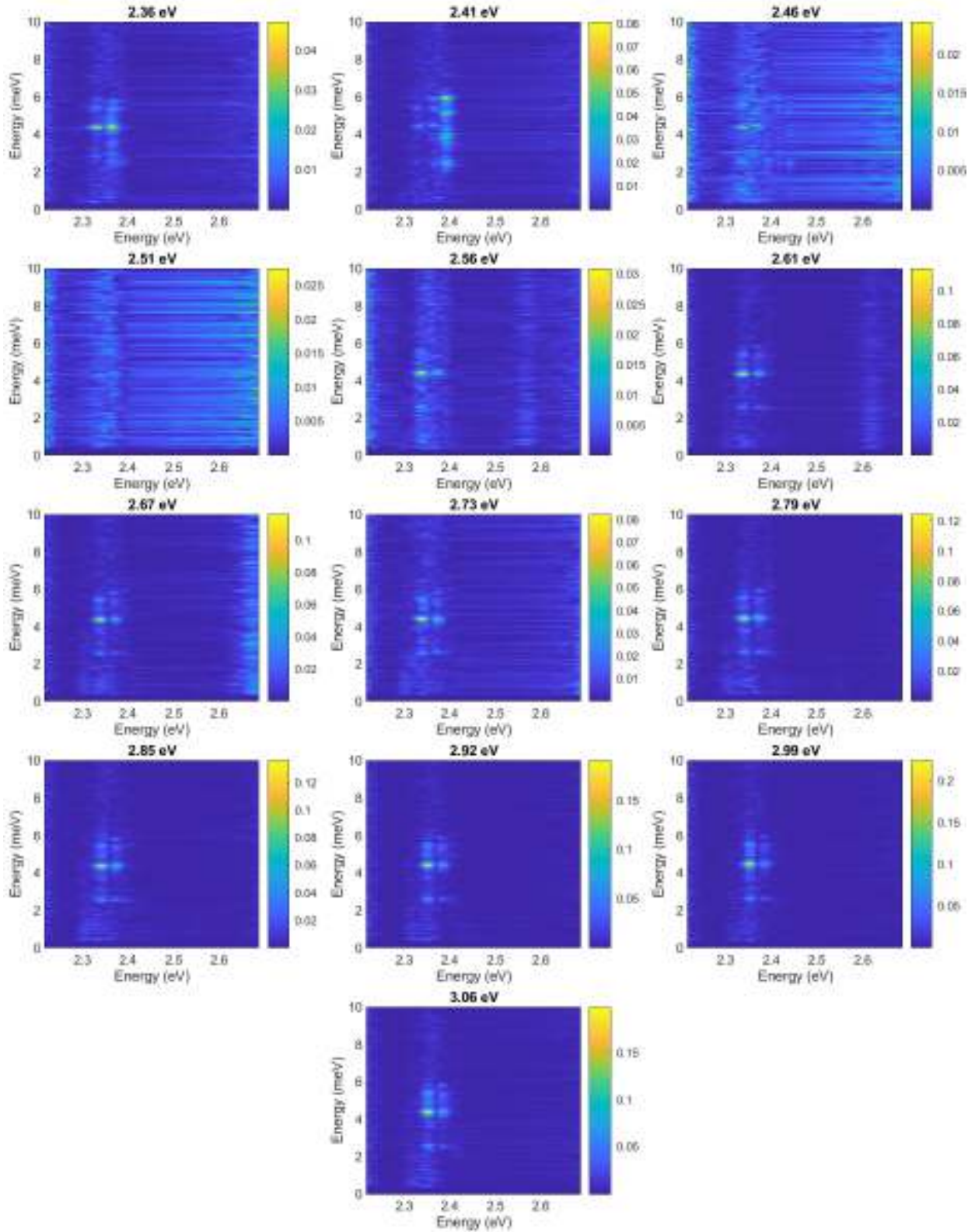

FIG. S4. Pump-energy dependent beating maps dataset from 2.36 eV to 3.06 eV. The pump energy is indicated in the title of each subfigure.

5

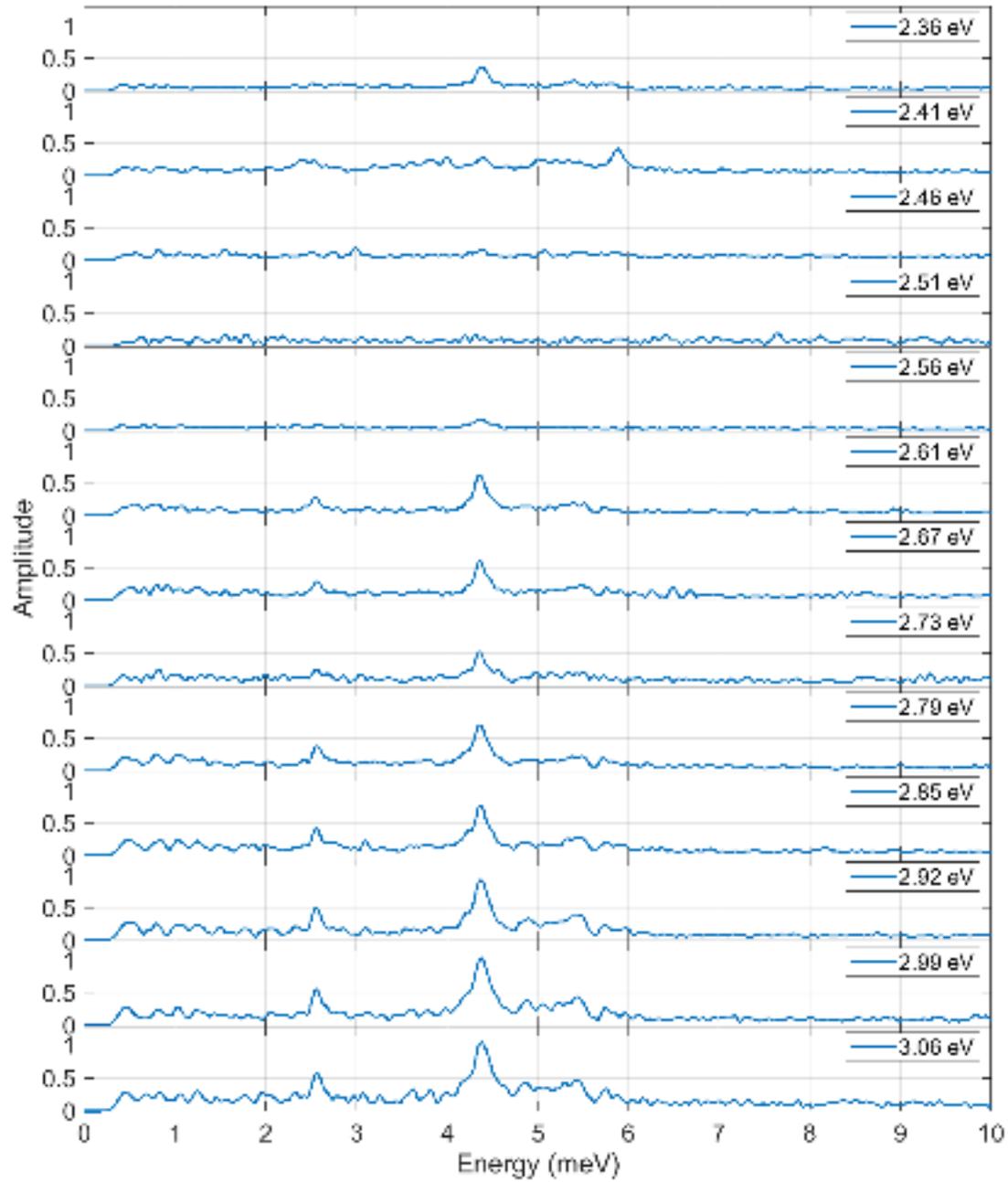

FIG. S5. Pump-energy dependence dataset from 2.36 eV to 3.06 eV, integrated along the probe axis.

6

## C. Full fluence dependence dataset on (PEA)$_2$PbI$_4$ at 5K

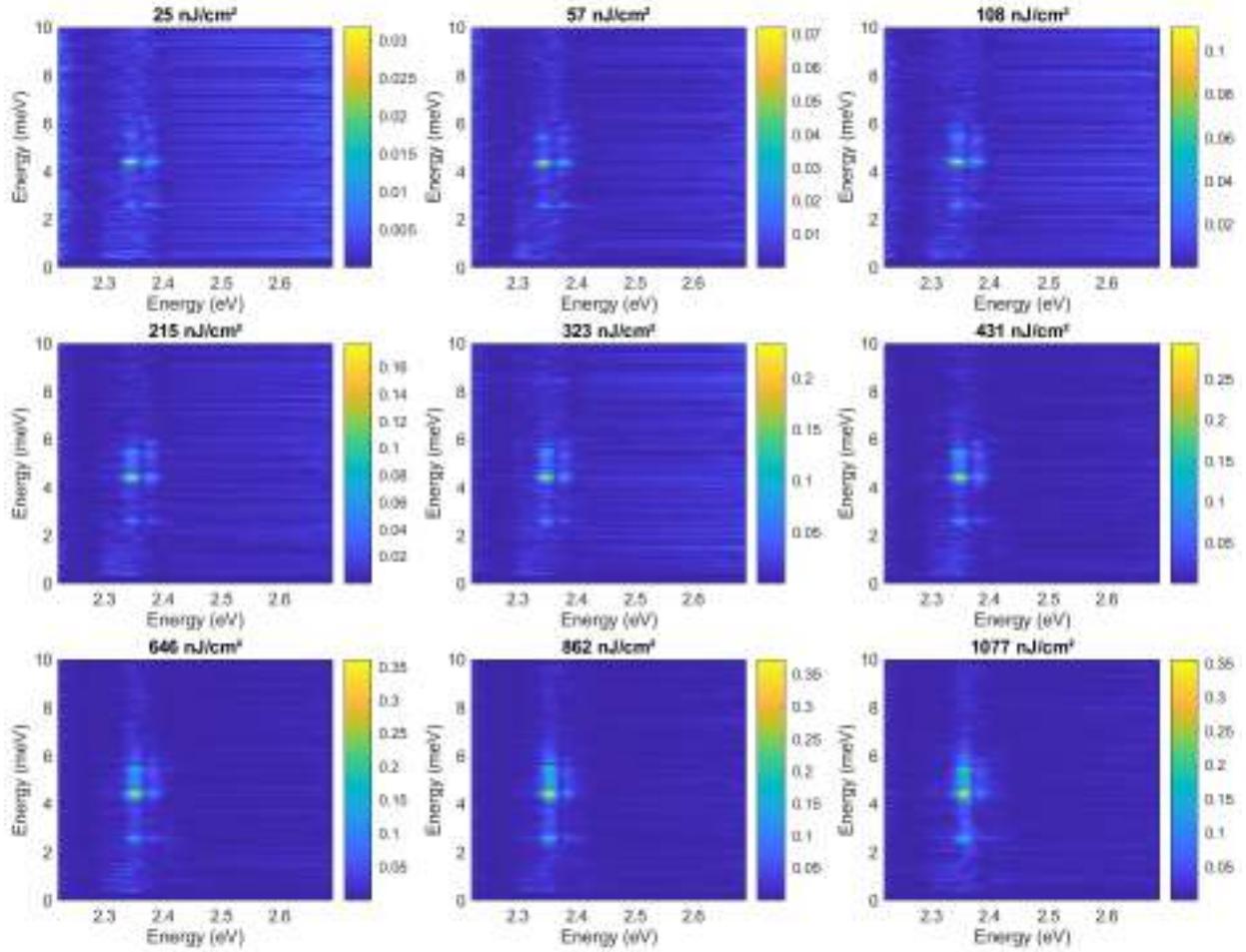

FIG. S6. Fluence dependent beating maps dataset. The fluence used for each experiment is indicated in the title of each subfigure.

7

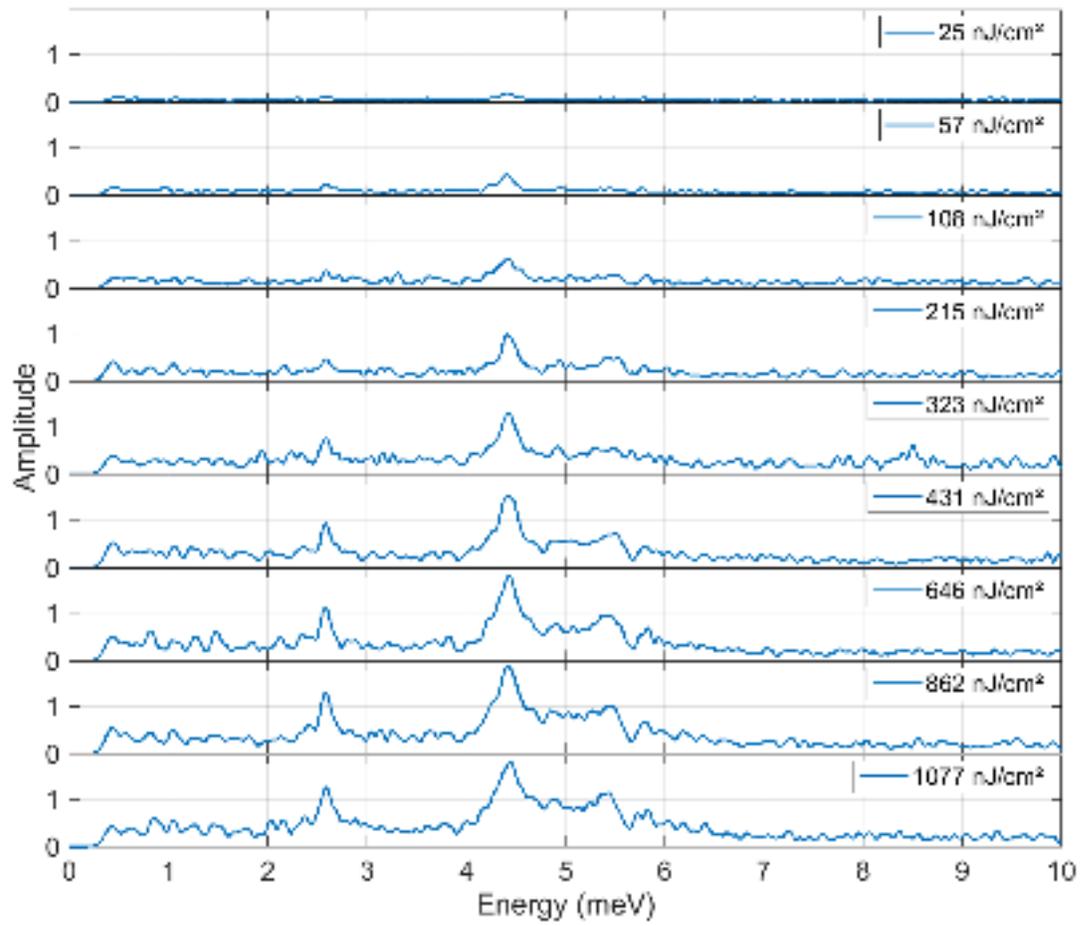

FIG. S7. Fluence dependence dataset, integrated along the probe axis.

### D. Resonant impulsive stimulated Raman spectroscopy on (NBT)$_2$PbI$_4$ at 5K

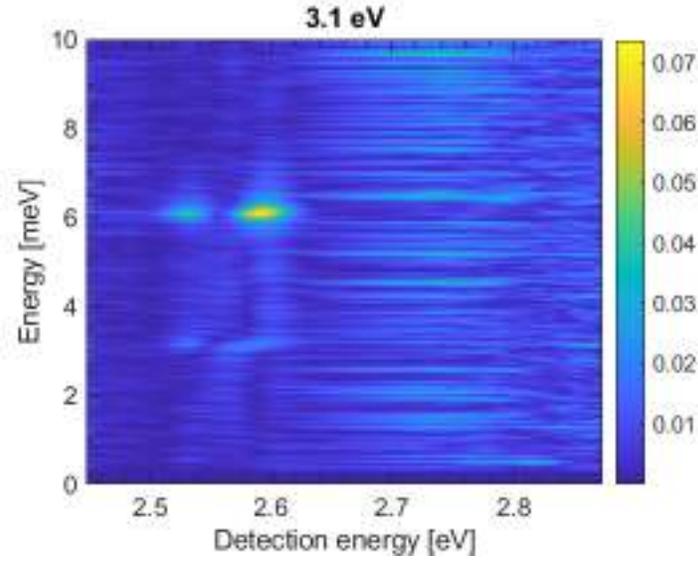

FIG. S8. Pump-energy dependent beating maps dataset pumped into the continuum band at 3.1 eV.

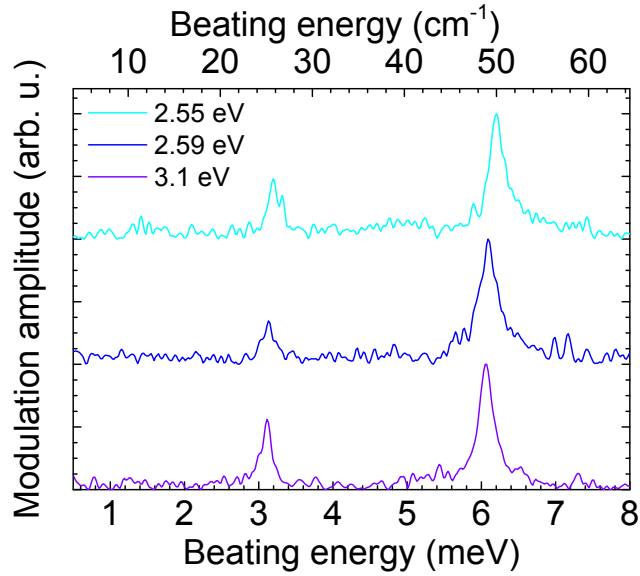

FIG. S9. Pump-energy dependence dataset integrated along the probe axis pumped at 2.55 eV, 2.59 eV and 3.1 eV.

### E. Supplementary transient absorption data for (PEA)$_2$PbI$_4$

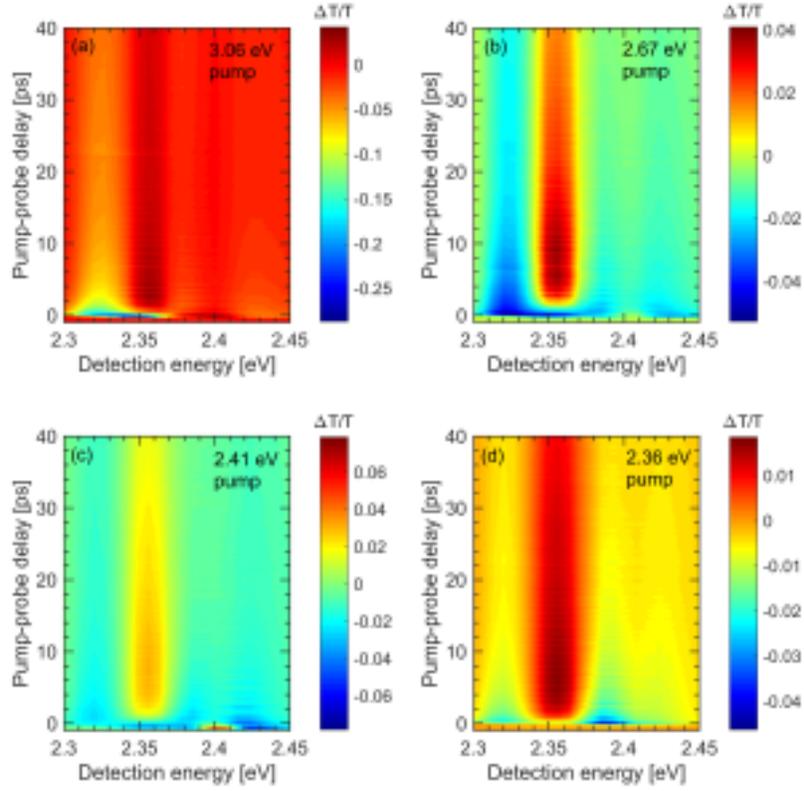

FIG. S10. Raw pump-probe signal for pump energies of 3.06 eV (a), 2.67 eV (b), 2.41 eV (c) and 2.36 eV (d). As mentioned in the main text, photon fluences were kept constant during the measurements.

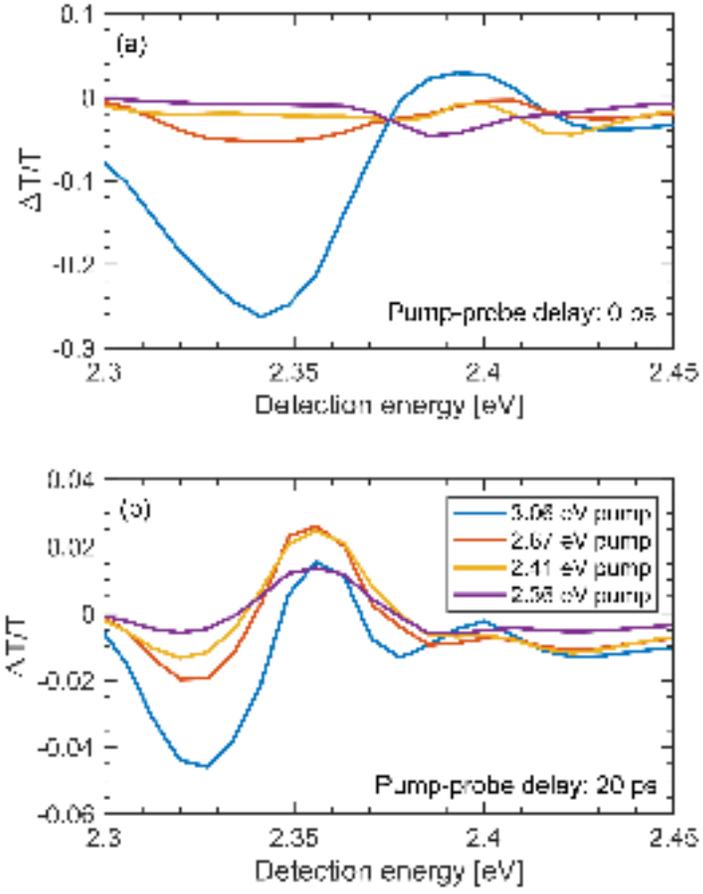

FIG. S11. Pump-probe signal for pump energies of 3.06 eV, 2.67 eV, 2.41 eV and 2.36 eV at pump-probe delays of 0 (a) and 20 ps (b).

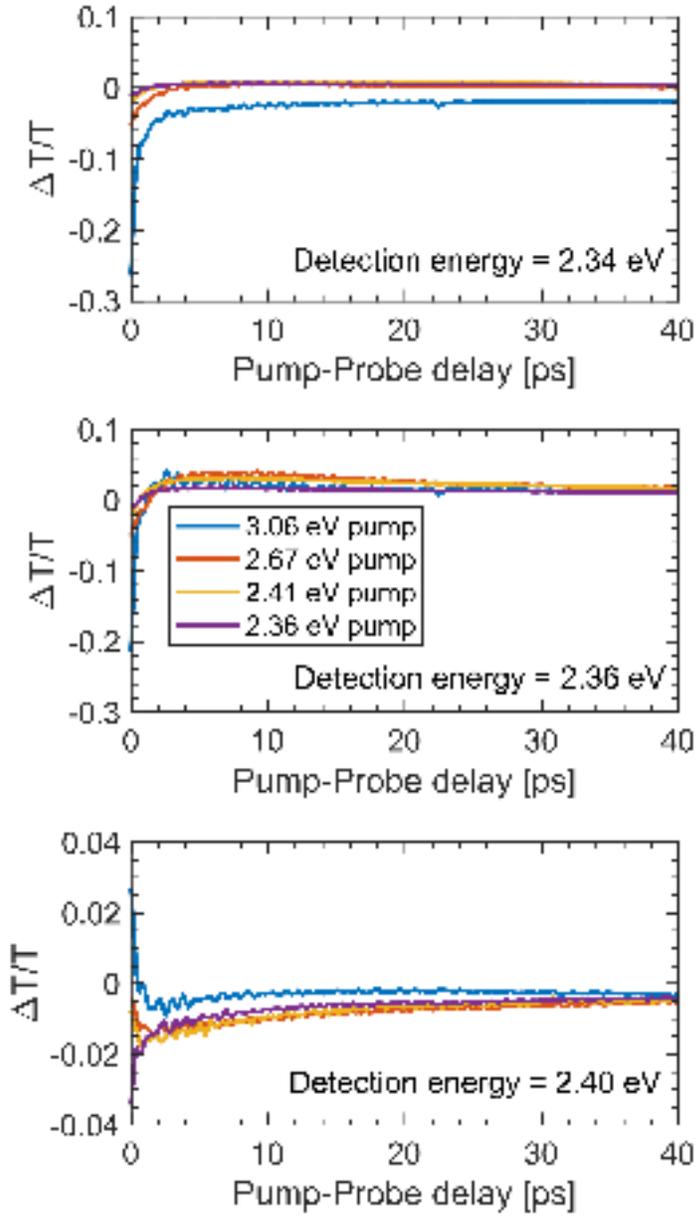

FIG. S12. Pump-probe dynamics for pump energies of 3.06 eV, 2.67 eV, 2.41 eV and 2.36 eV at detection energies of 2.34 eV (a), 2.36 eV (b) and 2.40 eV (c)

## F. Pump excitation spectra

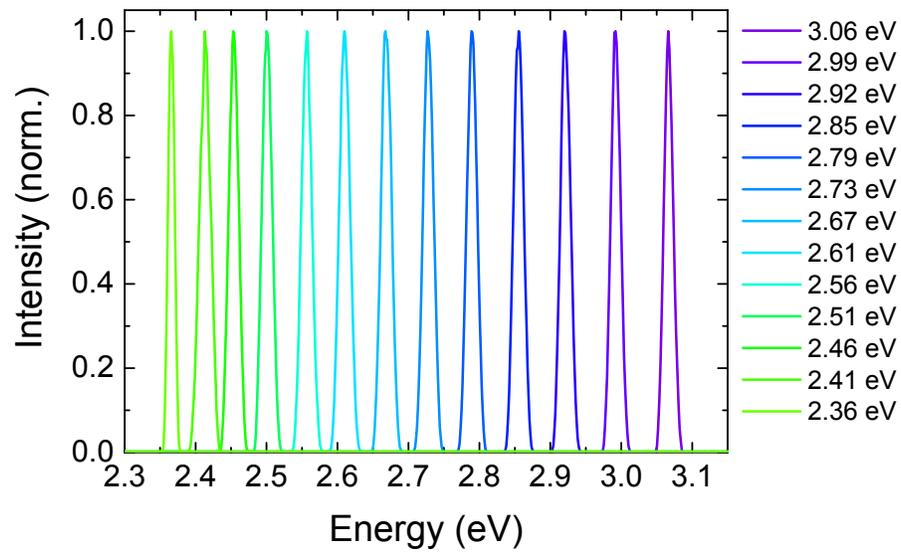

FIG. S13. Pump excitation spectra used to acquire spectra of the pump-energy dependent dataset

## S3. WAVEPACKET DYNAMICS SIMULATIONS

The detection energy dependence of the observed mode is very similar to that predicted by the modulation of the dielectric permittivity by vibrational wavepacket dynamics induced by the pump pulse. The symetric lineshape with a node at the peak of the electronic resonance accompanied by a phase-shift has been previously observed in many systems such as single walled semiconducting carbon nanotubes.[1]

To verify to which extent this simple model is applicable to this complex system, we follow the work of Kumar et al..[2] Their approach separates the density matrix into its electronic and vibrational degrees of freedom and computes an effective change in the equilibrium first-order permittivity induced by excitation by the pump pulse. This useful trick allows us to easily distinguish between pump and probe interactions while drastically reducing the required computation resources when compared to a full $\chi^{(3)}$ calculation.

The electronic density matrix is taken to be diagonal, neglecting effects from electronic coherences. This approximation is valid in the case of pump-probe measurements on two-level systems, but breaks down when more electronic levels are taken in consideration. The problem is then reduced to the computation of the vibrational degrees of freedoms effect on the dielectric permittivity.

To model these, the pump-induced modulation of the vibrational density matrices are computed for the ground and excited electronic states. The vibrational Hamiltonian is the same in both cases, except for a displacement $\Delta$ along the nuclear coordinate associated to the vibrational mode. When the pump field interacts with the vibrational part of the system, it imparts a displacement in real and momentum space modeled using a moment generating function. All pump interactions are contained within this initial kick, after which the vibrational degree of freedom evolve freely according to their respective Hamiltonian before interacting with the probe pulse.

The only free parameters of this model are the displacement of the vibrational Hamiltonian $\Delta$ upon excitation and the complex lineshape of the system under study. The latter supposes a single homogeneously broadened electronic state and its accompanying vibrational replica. With these considerations and supposing undamped wavepacket motion, analytical expressions for the amplitude of the pump-probe modulation can be derived, and further reduce the required computation time. To account for inhomogeneous broadening,

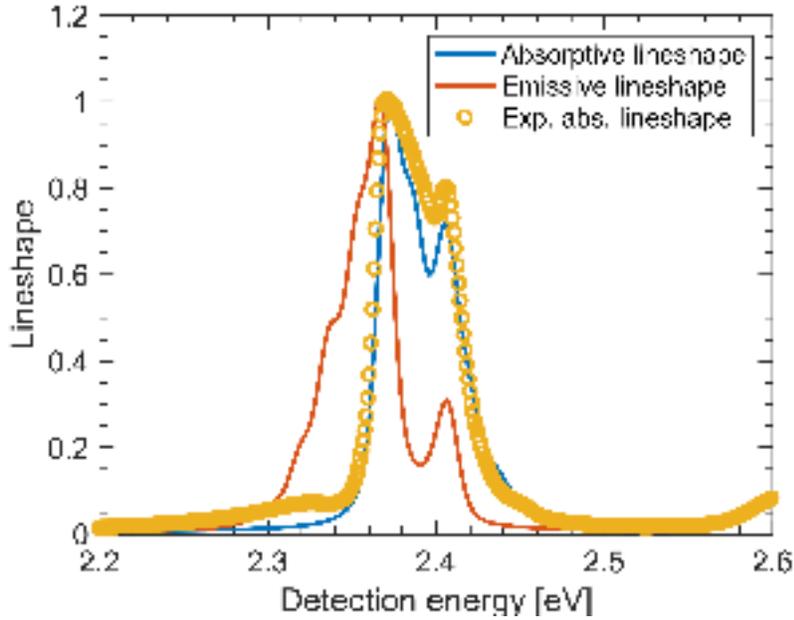

FIG. S14. Absorptive (blue) and emissive (red) lineshapes used in the wavepacket dynamics simulations. The experimental lineshape is also shown in yellow for comparison.

the signal is averaged across an Gaussian profile of given width, another free parameter in our model.

To obtain the complex lineshape from which stem these calculations, we use the displacements and relative electronic amplitudes obtained previously from the fit of a modified two-dimensional Eliotts formula.[3] Distinct electronic transitions are treated as inhomogeneous broadening and are not considered to be coherently coupled. This is a flaw of this simple model, as previous two-dimensional coherent excitation spectroscopy have shown these to be coherently coupled. As shown below in figure S14, the agreement between the obtained lineshapes and the absorption spectra is far from perfect, due to the different lineshapes used here (Lorentzians) and in the modified Eliotts formula (secant-hyperbolic functions).

The resulting simulations using these lineshapes are presented in figure S15. For panels (a) and (c), only the main peak and its vibrational replicas were used. The simulated phase and amplitude qualitatively reproduce some of the features of the experimental data when pumping above the conduction band , however, the exact phase profile and the spectrums asymmetry are not reproduced. For panels (b) and (d), the main resonance and its vibrational replicas was used as well as a secondary electronic transition 35 meV above. The

15

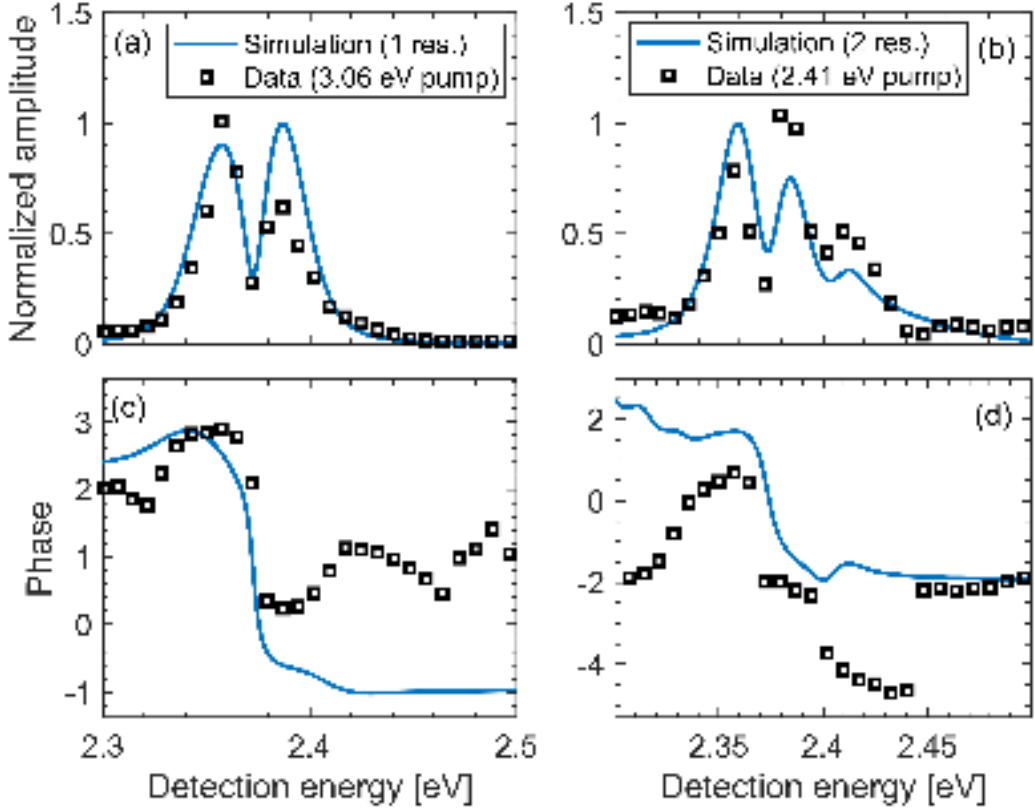

FIG. S15. Simulations showing the best agreement with experimental data. The amplitude ((a) and (b)) and phase ((c) and (d)) profiles when exciting in the conduction band ((a) and (c)) or the B-exciton ((b) and (d)) are shown alongside experimental data (black squares). For (a) and (c), only the A exciton's lineshape was considered while both excitons were considered in (b) and (d).

simulation is compared with the data obtained when pumping directly the blue end of the exciton spectrum. While the agreement between the simulations and the data is minimal, qualitative features are reproduced. The optimal displacements $\Delta$ along M2's general coordinate obtained by fitting this model to the experimental data are 1.32 for when pumping the conduction band and 1.07 and 0.47 for the A and B excitons respectively when pumping the B exciton.

## S4. FULL PHASE AND AMPLITUDE CUTS OF BEATING MAPS

The data presented in the main text shows the amplitude of the beating maps for a chosen set of modes under different pumping conditions. By nature of the Fourier transform, this

is only part of the system's response, the other one being the phase profile. The analysis is here extended to the phase profile, which shows jumps of about $\pi$ radians in the vicinity of an electronic resonance and further supports our conclusions.

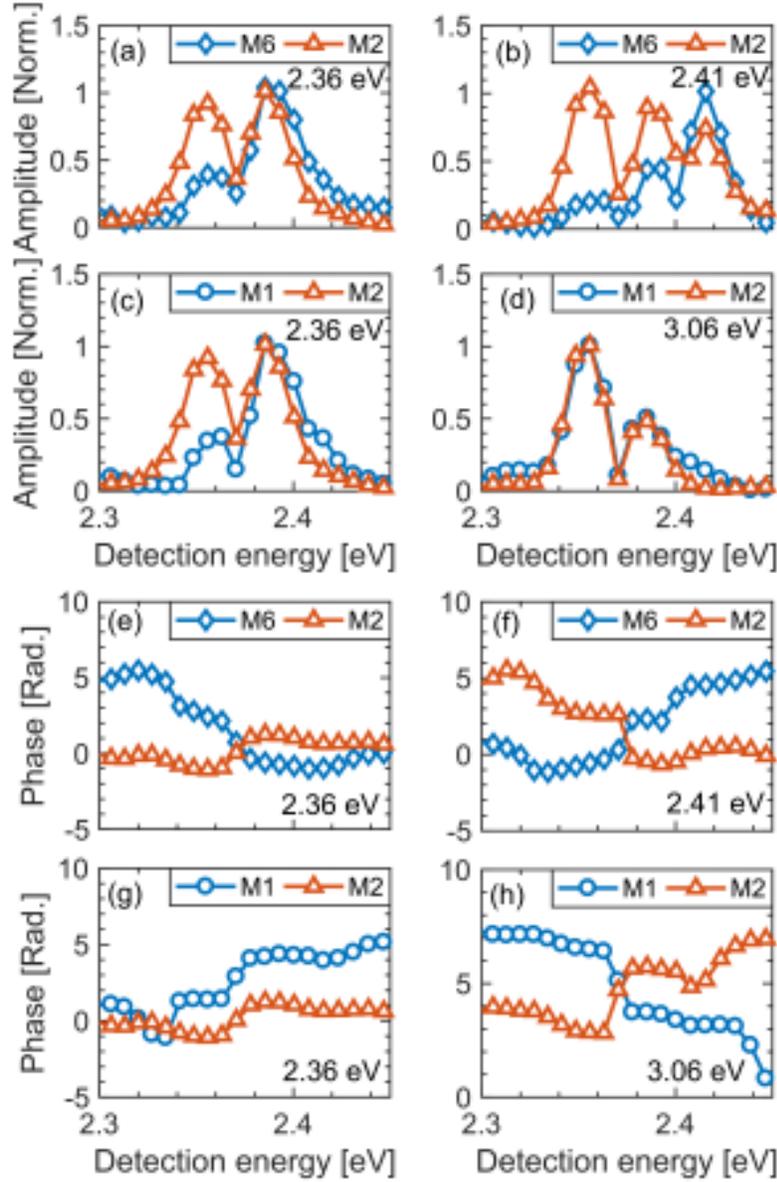

FIG. S16. Full phase and amplitude cuts of beating maps for various modes for pumping energies of 2.36 eV (a,c,e and g), 2.41 eV (b and f) and 3.06 eV (d and h). The amplitude and phase are presented in panels a to d and e to h respectively.

17

## S5. DFT CALCULATIONS

| MODE # | FREQ. (CM-1) | IR (KM/MOL) | I_tot | RAMAN I_par | I_perp | relax. en. (meV) |
|---|---|---|---|---|---|---|
| 1 | 0,01 | 0,0010 | 0,00 | 0,00 | 0,00 | 0,00 |
| 2 | 0,01 | 0,0011 | 0,00 | 0,00 | 0,00 | 0,00 |
| 3 | 0,02 | 0,0022 | 0,00 | 0,00 | 0,00 | 0,00 |
| 4 | 15,98 | 1,9817 | 0,00 | 2,77 | 1,59 | 1,18 | 0,02 |
| 5 | 18,57 | 2,3020 | 3,42 | 0,00 | 0,00 | 0,00 | 0,04 |
| 6 | 19,82 | 2,4574 | 2,59 | 0,00 | 0,00 | 0,00 | 0,01 |
| 7 | 23,98 | 2,9731 | 20,78 | 0,00 | 0,00 | 0,00 | 0,07 |
| 8 | 25,68 | 3,1837 | 0,00 | 1,15 | 0,79 | 0,36 | 0,61 |
| 9 | 29,45 | 3,6518 | 3,40 | 0,00 | 0,00 | 0,00 | 0,00 |
| 10 | 30,42 | 3,7717 | 3,03 | 0,01 | 0,01 | 0,00 | 0,01 |
| 11 | 31,70 | 3,9302 | 0,03 | 3,62 | 2,10 | 1,51 | 0,94 |
| 12 | 33,48 | 4,1509 | 2,79 | 0,15 | 0,09 | 0,06 | 0,01 |
| 13 | 33,92 | 4,2062 | 0,17 | 1,48 | 0,86 | 0,62 | 0,59 |
| 14 | 36,39 | 4,5121 | 4,55 | 4,78 | 3,09 | 1,70 | 5,21 |
| 15 | 36,41 | 4,5143 | 12,66 | 1,69 | 1,08 | 0,60 | 3,49 |
| 16 | 36,97 | 4,5837 | 2,85 | 0,02 | 0,01 | 0,01 | 0,16 |
| 17 | 41,38 | 5,1309 | 2,37 | 2,02 | 1,38 | 0,64 | 0,96 |
| 18 | 41,51 | 5,1464 | 2,85 | 1,92 | 1,31 | 0,61 | 0,73 |
| 19 | 43,07 | 5,3409 | 0,01 | 6,39 | 3,99 | 2,40 | 4,55 |
| 20 | 46,08 | 5,7138 | 31,03 | 0,00 | 0,00 | 0,00 | 0,00 |
| 21 | 47,29 | 5,8639 | 0,02 | 4,10 | 3,40 | 0,70 | 4,79 |
| 22 | 51,72 | 6,4135 | 29,71 | 0,00 | 0,00 | 0,00 | 0,25 |
| 23 | 53,33 | 6,6129 | 0,00 | 3,98 | 2,48 | 1,51 | 0,84 |
| 24 | 54,86 | 6,8021 | 0,08 | 18,72 | 10,79 | 7,94 | 0,25 |
| 25 | 57,25 | 7,0982 | 27,66 | 0,00 | 0,00 | 0,00 | 0,06 |
| 26 | 58,26 | 7,2235 | 51,53 | 0,01 | 0,01 | 0,00 | 0,01 |
| 27 | 59,05 | 7,3222 | 0,26 | 2,77 | 2,06 | 0,71 | 0,65 |
| 28 | 63,35 | 7,8548 | 174,12 | 0,00 | 0,00 | 0,00 | |
| 29 | 64,16 | 7,9559 | 0,17 | 8,60 | 5,10 | 3,50 | |
| 30 | 64,29 | 7,9717 | 46,04 | 0,03 | 0,02 | 0,01 | |
| 31 | 68,13 | 8,4470 | 9,03 | 0,00 | 0,00 | 0,00 | |
| 32 | 71,42 | 8,8556 | 259,64 | 0,02 | 0,02 | 0,01 | |
| 33 | 71,93 | 8,9189 | 0,62 | 9,98 | 7,52 | 2,46 | |
| 34 | 74,73 | 9,2654 | 0,10 | 15,40 | 9,20 | 6,20 | |
| 35 | 76,31 | 9,4616 | 15,92 | 0,01 | 0,01 | 0,01 | |
| 36 | 77,49 | 9,6076 | 0,03 | 2,13 | 1,50 | 0,63 | |
| 37 | 77,72 | 9,6373 | 35,05 | 0,01 | 0,01 | 0,00 | |
| 38 | 79,58 | 9,8672 | 41,56 | 0,00 | 0,00 | 0,00 | |
| 39 | 82,89 | 10,2777 | 60,44 | 0,04 | 0,03 | 0,01 | |
| 40 | 83,81 | 10,3916 | 0,16 | 13,53 | 9,36 | 4,18 | |
| 41 | 85,99 | 10,6625 | 0,01 | 7,23 | 4,45 | 2,78 | |
| 42 | 87,27 | 10,8208 | 3,71 | 0,01 | 0,01 | 0,00 | |
| 43 | 91,26 | 11,3156 | 168,07 | 0,00 | 0,00 | 0,00 | |
| 44 | 92,55 | 11,4759 | 0,00 | 10,13 | 6,42 | 3,70 | |
| 45 | 93,50 | 11,5936 | 0,04 | 9,36 | 5,38 | 3,97 | |
| 46 | 94,66 | 11,7370 | 32,32 | 0,01 | 0,00 | 0,00 | |
| 47 | 97,32 | 12,0671 | 0,01 | 3,48 | 1,99 | 1,49 | |
| 48 | 101,51 | 12,5865 | 0,23 | 0,00 | 0,00 | 0,00 | |

FIG. S17. Table of all vibrational modes obtained by DFT calculations showing their energy, Raman and infrared activities. The values tabulated in the main text are highlighted.

A.  DFT assignment of the observed vibrational signatures as hole-/electron-polaron

TABLE S1. **Hole-phonon and electron-phonon coupling parameters, computed for (PEA)$_2$PbI$_4$ and (NBT)$_2$PbI$_4$, for the vibrational modes in Tables I and II in the main manuscript.** Disentangled contributions of lattice relaxation around the hole- and the electron- for the vibrational normal modes of interest is obtained starting from Eq. 1 in the main manuscript. This procedure consists in tracking the variation of the band gap, resulting from the displacement of the atoms along the vibrational normal mode coordinates, as due to a shift in the valence or conduction band edge, respectively. To circumvent the fact that orbital energies are ill-defined quantities for periodic systems, we estimate the variation of the frontier orbital energies with respect to an electronic level, assumed fixed during the vibration (the 1s state of carbon, in the present case).

| Mode | Frequency [meV] | Electron phonon coupling Valence | Electron phonon coupling Conduction |
|---|---|---|---|
| (PEA)$_2$PbI$_4$ | | | |
| M1  | 3.18 | 0.06  | 1.76 |
| M1' | 3.93 | 1.87  | 0.02 |
| M2  | 4.51 | 17.78 | 2.37 |
| M3  | 4.51 | 0.01  | 4.42 |
| M5  | 5.34 | 4.33  | 0.06 |
| M6  | 5.86 | 1.73  | 0.90 |
| (NBT)$_2$PbI$_4$ | | | |
| N1 | 2.93 | 5.04  | 0.92 |
| N2 | 5.38 | 4.24  | 0.07 |
| N3 | 5.97 | 25.57 | 0.18 |
| N4 | 6.60 | 11.48 | 0.17 |
| N5 | 7.37 | 1.26  | 0.64 |

## B. Comparison of relaxation energies with and without Spin-Orbit Coupling (SOC)

TABLE S2. **Electron-lattice relaxation energy $\lambda$ (see Eq. 1 of main text) computed for $(PEA)_2PbI_4$ with and without spin-orbit coupling (SOC).**

| Mode | Frequency [cm$^{-1}$] | $\lambda$ [meV] NO SOC | $\lambda$ [meV] SOC |
|---|---|---|---|
| M1  | 2.57 | 0.61 | 0.97 |
| M1' | 4.38 | 0.93 | 0.98 |
| M2  | 4.89 | 5.16 | 4.45 |
| M3  | 5.22 | 3.46 | 1.74 |
| M5  | 5.41 | 4.55 | 1.61 |
| M6  | 5.75 | 4.80 | 6.07 |

## C. Animated diagrams of atomic motions involved in relevant vibratonnal modes

Animated files corresponding to the atomic motion of modes N1 to N5 and M1 to M6 are available for download as supplemental files. The templating cation and represented mode are specified in the file name. For instance, `NBT_vib_mode_N1.mov` corresponds to motion of mode N1 in $(NBT)_2PbI_4$.

---

[1] Lüer, L. *et al.* Coherent phonon dynamics in semiconducting carbon nanotubes: A quantitative study of electron-phonon coupling. *Phys. Rev. Lett.* **102**, 127401 (2009).

[2] Kumar, A. T., Rosca, F., Widom, A. & Champion, P. M. Investigations of amplitude and phase excitation profiles in femtosecond coherence spectroscopy. *J. Chem. Phys.* **114**, 701–724 (2001).

[3] Neutzner, S., Cortecchia, D. & Petrozza, A. Exciton-polaron spectral structures in two dimensional hybrid lead-halide perovskites. *Phys. Rev. Materials* **2**, 064605 (2018).



\end{document}